\documentclass[%
 reprint,
 superscriptaddress,
nofootinbib,
 amsmath,amssymb,
 aps,
 prx,
]{revtex4-2}

\usepackage{graphicx}
\usepackage{dcolumn}
\usepackage{bm}
\usepackage{multirow}
\usepackage[colorlinks=true, citecolor=blue, urlcolor=blue, linkcolor=blue, breaklinks=true, pdfpagelabels=false]{hyperref}
\usepackage{algpseudocode}
\usepackage{amsmath, amssymb}
\usepackage{booktabs}
\usepackage{color}
\newcounter{alg}
\renewcommand{\thealg}{\arabic{alg}}
\newcommand{\algcaption}[1]{%
  \refstepcounter{alg}%
  \textbf{Algorithm~\thealg.}~#1%
}

\newenvironment{ruledalgorithm}[1][\columnwidth]{%
  \par\addvspace{6pt}%
  \noindent\hrule height 0.8pt\relax
  \vspace{2pt}%
  \noindent\begin{minipage}{#1}%
    \setlength{\parskip}{4pt}%
    \setlength{\parindent}{0pt}%
}{%
  \end{minipage}%
  \vspace{2pt}%
  \noindent\hrule height 0.8pt\relax
  \par\addvspace{6pt}%
}

\def\figureautorefname~#1\null{Fig.\,#1\null}
\def\equationautorefname~#1\null{Eq.\,(#1)\null}

\begin{document}

\title{Revisiting the broken symmetry phase of solid hydrogen: A neural network variational Monte Carlo study}

\author{Shengdu Chai}
\affiliation{Interdisciplinary Center for Theoretical Physics and Information Sciences (ICTPIS), Fudan University, Shanghai 200433, China}
\affiliation{Shanghai Artificial Intelligence Laboratory, Shanghai 200232, China}
\author{Chen Lin}
\affiliation{Department of Engineering Science, University of Oxford, Oxford OX1 3PJ, United Kingdom}
\author{Xinyang Dong}
\affiliation{Beijing National Laboratory for Condensed Matter Physics and Institute of Physics,
  Chinese Academy of Sciences, Beijing 100190, China}
\author{Yuqiang Li}
\affiliation{Shanghai Artificial Intelligence Laboratory, Shanghai 200232, China}
\author{Wanli Ouyang}
\affiliation{Shanghai Artificial Intelligence Laboratory, Shanghai 200232, China}
\affiliation{Department of Information Engineering, The Chinese University of Hong Kong, Hong Kong SAR  HKG, China}
\author{Lei Wang}
\affiliation{Beijing National Laboratory for Condensed Matter Physics and Institute of Physics,
  Chinese Academy of Sciences, Beijing 100190, China}
\author{X. C. Xie}
\affiliation{Interdisciplinary Center for Theoretical Physics and Information Sciences (ICTPIS), Fudan University, Shanghai 200433, China}
\affiliation{International Center for Quantum Materials, School of Physics, Peking University, Beijing 100871, China}
\affiliation{Hefei National Laboratory, Hefei 230088, China}

\date{\today}

\begin{abstract}
  The crystal structure of high-pressure solid hydrogen remains a fundamental open problem. Although the research frontier has mostly shifted toward ultra-high pressure phases above 400 GPa, we show that even the broken symmetry phase observed around 130~GPa requires revisiting due to its intricate coupling of electronic and nuclear degrees of freedom. Here, we develop a first principle quantum Monte Carlo framework based on a deep neural network wave function that treats both electrons and nuclei quantum mechanically within the constant pressure ensemble. Our calculations reveal an unreported ground-state structure candidate for the broken symmetry phase with $Cmcm$ space group symmetry, and we test its stability up to 96 atoms. The predicted structure quantitatively matches the experimental equation of state and gives the closest x-ray diffraction peak-position match among the tested candidates. Furthermore, our group-theoretical analysis provides a symmetry-counting compatibility check between the $Cmcm$ structure and existing Raman and infrared spectroscopic data. Crucially, static density functional theory calculation reveals the $Cmcm$ structure as a dynamically unstable saddle point on the Born-Oppenheimer potential energy surface, demonstrating that a full quantum many-body treatment of the problem is necessary. These results shed new light on the phase diagram of high-pressure hydrogen and call for further experimental verifications.
\end{abstract}
\maketitle

\section{Introduction}

Understanding the behavior of hydrogen under extreme pressure remains a grand challenge with profound implications~\cite{Guillot_2005,mcmahon_propertieshydrogen_2012,gregoryanz_everythingyou_2020,wigner_possibility_1935,PhysRevLett.21.1748}, ranging from planetary interiors to the pursuit of the metallic, high-temperature superconducting state. Determining the precise structure is fundamental to these goals, as structure dictates the physical properties. Although decades of experimental effort have revealed a rich phase diagram, the detailed structures of high-pressure hydrogen remain largely unsettled. To date, only the structure of low-pressure Phase I is firmly confirmed by experiment as a hexagonal close-packed (hcp) lattice of freely rotating $\mathrm{H}_2$ molecules~\cite{hazen_singlecrystalxray_1987,mao_synchrotronxray_1988,hemley_equationstate_1990}. With increasing pressure, the system transitions into a broken symmetry phase characterized by the emergence of long-range orientational order. Early spectroscopic observations established this transition in para-H$_2$, D$_2$, and HD~\cite{lorenzana_orientationalphase_1990,silvera_newlowtemperature_1981,moshary_remarkable_1993}. While recent frontiers have advanced toward the ultra-high pressure regime ($>$400 GPa) in pursuit of metallic hydrogen, the detailed structures of the broken symmetry phase are less settled than commonly assumed and require revisiting.

Despite tremendous progress in experimental techniques~\cite{hazen_singlecrystalxray_1987,mao_synchrotronxray_1988,hemley_equationstate_1990,loubeyre_compression_2022,akahama_ramanscattering_2017,goncharov_newhighpressure_1998,goncharov_vibronfrequencies_2011,goncharov_spectroscopicstudies_2001,hemley_phasetransition_1988,hemley_spectroscopicstudies_1998,liu_highpressurebehavior_2017,lorenzana_orientationalphase_1990,mazin_quantumclassical_1997,hemley_lowfrequency_1990,ji_ultrahighpressureisostructural_2019,ji_ultrahighpressurecrystallographic_2025,akahama_evidencexray_2010,goncharenko_neutronxray_2005,loubeyre_xraydiffraction_1996}, it is still difficult to unambiguously determine the structure of solid hydrogen under extreme conditions. The extreme pressure conditions and the small, fragile samples restrict feasible probes to low-power optical techniques, such as Raman and infrared (IR) spectroscopy~\cite{loubeyre_compression_2022,akahama_ramanscattering_2017,goncharov_newhighpressure_1998,goncharov_vibronfrequencies_2011,goncharov_spectroscopicstudies_2001,hemley_phasetransition_1988,hemley_spectroscopicstudies_1998,liu_highpressurebehavior_2017,lorenzana_orientationalphase_1990,mazin_quantumclassical_1997,hemley_lowfrequency_1990}. Moreover, the inherently weak x-ray diffraction (XRD) signals of hydrogen~\cite{ji_ultrahighpressureisostructural_2019,ji_ultrahighpressurecrystallographic_2025,akahama_evidencexray_2010,goncharenko_neutronxray_2005,loubeyre_xraydiffraction_1996} further hinder direct structural identification, making it difficult to discern the signatures of the high-pressure phases.

In light of these experimental difficulties, first principles calculations have proposed several candidate structures for broken symmetry phase, including $P2_1/c\text{-}8$, $P2_1/c\text{-}24$, $Pca2_1$, $Cmc2_1$, $P6_3/m$, \textit{etc}.~\cite{kohanoff_solid_1997,nagao_initio_1999,kitamura_quantumdistribution_2000,stadele_metallizationmolecular_2000,johnsonStructureBandgapClosure2000,zhang_initio_2006,pickard_structurephase_2007,zhang_brokensymmetry_2007,pickard_structures_2009,moraldi_orientational_2009,li_classicalquantum_2013,azadi_fatedensity_2013,azadi_quantummonte_2013,drummond_quantummonte_2015,zong_understandinghigh_2020,hellgren_highpressureiiiii_2022,li_highpressurestructures_2024}, and others even suggested that broken symmetry phase to be a ``quantum fluxional solid"~\cite{biermann_quantumeffects_1998,geneste_strongisotope_2012}. However, these predictions have been inconclusive due to the inherent limitations of standard computational frameworks rooted in the Born-Oppenheimer (BO) approximation and the inadequate treatment of electronic correlations. Static Density Functional Theory (DFT) calculations~\cite{Bonitz:2024ags,monserrat_hexagonalstructure_2016,morales_predictivefirstprinciples_2013,nagara_stablephases_1992,surh_zeropoint_1993} neglect the zero-point energy (ZPE) of protons, a contribution comparable to the cohesive energy in hydrogen, often leading to erroneous stability rankings. While~\textit{ab initio} molecular dynamics (AIMD)~\cite{ackland_structuressolid_2020,Bonitz:2024ags,kohanoff_dipolequadrupoleinteractions_1999} introduces thermal effects, it typically treats nuclei as classical particles, thereby missing critical nuclear quantum effects (NQEs)~\cite{markland_nuclearquantum_2018}, such as tunneling and zero-point motion. Consequently, classical AIMD cannot intrinsically capture the distinct phase behaviors of isotopes. More advanced methods like path-integral molecular dynamics~\cite{Bonitz:2024ags,niu_stablesolid_2023} or path-integral quantum Monte Carlo ~\cite{Bonitz:2024ags,cui_rotationalordering_1997,kaxiras_orientationalorder_1994,surh_initiocalculations_1997} rigorously incorporate NQEs and recover isotope effects; however, they rely on an underlying potential energy surface generated by DFT, thus inheriting the systematic errors of approximate exchange-correlation functionals.

Crucially, the contribution of lattice anharmonicity, often neglected or treated insufficiently, is pivotal in determining the structural stability of solid hydrogen~\cite{azadi_dissociation_2014, Monacelli:2022cex}. Previous studies have shown that anharmonic correction is ``small but significant" and impacts molecular and atomic phases oppositely, significantly shifting predicted phase boundaries~\cite{azadi_dissociation_2014}. Even state-of-the-art approaches designed to capture these effects, such as the stochastic self-consistent harmonic approximation (SSCHA)~\cite{Monacelli:2022cex}, remain constrained by the BO approximation and the assumption of a Gaussian nuclear density matrix. To accurately describe the phase diagram, it is therefore essential to transcend these limitations by employing a framework that supports an intrinsically non-Gaussian nuclear wave function, provides a highly accurate description of electronic correlations, and treats electrons and nuclei on an equal quantum footing to recover nonadiabatic effects.

Neural network based variational Monte Carlo calculation (NNVMC), pioneered by~\cite{carleo_solving_2017} and in particular Ref.~\cite{pfau_abinitio_2020} for \textit{ab initio} simulations, has emerged as a powerful way to solve quantum many-body problems accurately and reliably~\cite{hermann_initio_2023,qian_deep_2025}.
To date, there have been applications to a broad set of physical and chemical problems, including molecules~\cite{han_solving_2019, pfau_abinitio_2020,Hermann:2020xqs,Pfau:2023azx,foster_initio_2025,li_spinadapted_2025} and solid-state materials~\cite{li_initio_2022}, moiré systems~\cite{li_emergent_2024,luo_simulating_2024}, homogeneous electron gases~\cite{Cassella:2022boh, pescia_messagepassing_2024, smith_unified_2024, valenti_critical_2025, ge_visualizing_2025,li_attention_2025}, excitons~\cite{Luo:2023tha}, superfluids~\cite{lou_neural_2024}, high-pressure hydrogen~\cite{xie_deep_2023a, dong_deep_2025, linteau_universal_2025}, fractional quantum Hall states~\cite{qian_describing_2025,teng_solving_2025, gattu_dressing_2025a, abouelkomsan_topological_2025a}, and topological insulators~\cite{li_deep_2025}. Collectively, these studies establish neural-network quantum states as a versatile and systematically improvable framework for first-principle quantum many-body computation.

Building on this progress, we develop a beyond-Born-Oppenheimer real-space neural network framework for enthalpy extremization to investigate solid hydrogen structures. By extending the NNVMC method to the constant pressure (NPT) ensemble~\cite{Linteau:2024gpe}, our framework simultaneously optimizes the crystal geometry and the full quantum many-body wave function for both electrons and nuclei~\cite{wang_full_2025}. Akin to the approach demonstrated in Refs.~\cite{linteau_universal_2025,Linteau:2024gpe}, our framework employs a highly flexible, non-Gaussian representation of the nuclear wave function, thereby treating NQEs and electronic correlations on an equal footing without relying on the BO approximation under NPT ensemble.

In this paper, we apply this methodology to provide new insights into the long-standing issue of the broken symmetry phase of solid hydrogen. Our simulations identify a candidate orthorhombic structure with $Cmcm$ space group symmetry. This candidate is compatible with available experimental observations and gives the closest XRD peak-position match among the tested candidates after uniform volume rescaling. Furthermore, symmetry analysis shows that its allowed vibrational modes are compatible with the spectroscopic signatures observed in Raman and IR experiments.

The remainder of the paper is organized as follows: Section~\ref{sec:method} introduces the computational framework and the neural network ansatz; Section~\ref{sec:result} presents the main results, including benchmark results and the newly identified $Cmcm$ structure with its experimental signatures, and demonstrates through static DFT analysis the importance of a fully quantum many-body treatment of the problem; Section~\ref{sec:conclusion} concludes with a discussion of the implications for the high-pressure hydrogen phase diagram and future directions.

\section{Methods}
\label{sec:method}

\begin{figure*}
  \includegraphics[width=\linewidth]{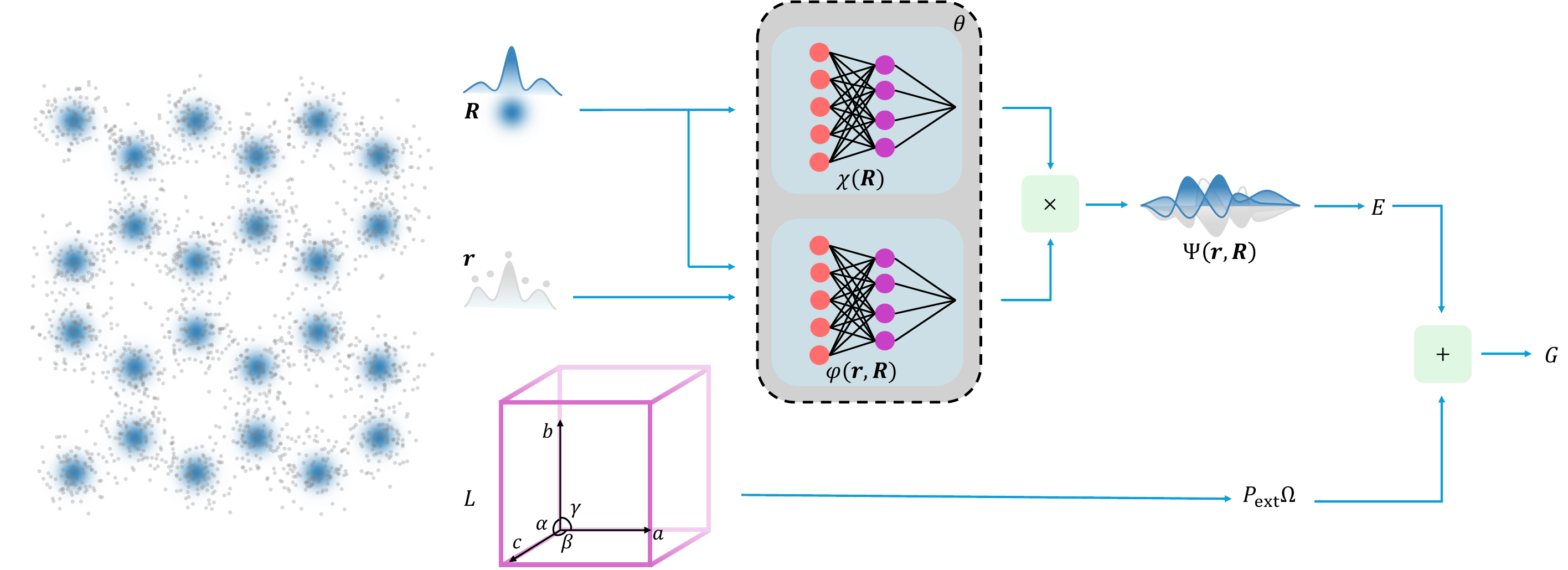}
  \caption{ The computational framework.
    The nuclear coordinates $\mathbf{R}$ and the electron coordinates $\mathbf{r}$ are fed into two neural networks that generate the nuclear wave function $\chi(\mathbf{R})$ and the electronic wave function $\varphi(\mathbf{r},\mathbf{R})$, respectively. Their product, $\Psi(\mathbf{r},\mathbf{R}) = \chi(\mathbf{R})\,\varphi(\mathbf{r},\mathbf{R})$, defines the total trial wave function. Together with the lattice parameters, they determine the enthalpy of the system $G = E + P_{\mathrm{ext}}\Omega$, where $P_{\mathrm{ext}}$ is the external pressure and $\Omega$ the cell volume. Treating the
    enthalpy as the objective function, the lattice parameters $L$ are optimized via simulated annealing, whereas the neural-network parameters $\theta$ are optimized by gradient descent.
  }
  \label{fig:framework}
\end{figure*}

We solve the ground state of the full Schr\"odinger equation for a system of $N$ hydrogen atoms in a periodic simulation cell with constant pressure. The Hamiltonian is
\begin{equation}
  \hat{H} = -\frac{1}{2}\sum_{i=1}^{N}\nabla_{\mathbf{r}_i}^2 - \frac{1}{2M}\sum_{I=1}^{N}\nabla_{\mathbf{R}_I}^2 + V(\mathbf{r}, \mathbf{R}),
  \label{eq:hamiltonian}
\end{equation}
where $\mathbf{R}=\{\mathbf{R}_I\}$ and $\mathbf{r}=\{\mathbf{r}_i\}$ denote the proton and electron positions, respectively. The first two terms denote the electron and proton kinetic energy, respectively. The last term represents the Coulomb interaction, calculated with the Ewald summation technique. We adopt atomic units ($\hbar=e=m_e=1$), consequently, energies are expressed in Hartree, lengths in Bohr, and the proton mass is set to $M=1836$. The overall framework is illustrated in Fig.~\ref{fig:framework}.

\subsection{Wave function ansatz}

Building a physically valid and expressive wave function is central to the success of the NNVMC framework.
We parametrize the ground state wave function as a product of two components
\begin{equation}
  \Psi(\mathbf{r},\mathbf{R}) \;=\; \chi(\mathbf{R}) \,\varphi(\mathbf{r},\mathbf{R}),
  \label{eq:wavefunction}
\end{equation}
where we denote $\chi(\mathbf{R})$ as the nuclear wave function and $\varphi(\mathbf{r},\mathbf{R})$ as the electronic wave function. Note that Equation~(\ref{eq:wavefunction}) is different from the Born-Huang expansion~\cite{born_dynamical_1955} in which $\varphi(\mathbf{r},\mathbf{R})$ is treated as electronic eigenstates given nuclear positions. On the other hand, we do not deem $\varphi(\mathbf{r},\mathbf{R})$ to be normalized with respect to the electron coordinates either. Hence, such an ansatz is not the same as the exact factorization ansatz~\cite{abedi_exact_2010,hunter_conditionalprobability_1975}. Equation~(\ref{eq:wavefunction}) is just a convenient factorization for the ground state wave function of the electron-nuclei coupled system in Eq.~(\ref{eq:hamiltonian}), designed to capture the coupled electron-nuclear quantum dynamics without the BO approximation.

We discuss the symmetry requirement of the wave functions and their neural network parametrizations in the following. $\varphi(\mathbf{r},\mathbf{R})$ must be antisymmetric under exchange of electrons and obey periodic boundary conditions (PBCs).
$\chi(\mathbf{R})$ must obey PBCs as well, while nuclear exchange symmetry is neglected based on nuclear localization in the molecular broken symmetry phase and prior QMC results showing negligible proton antisymmetrization in a rotationally ordered molecular phase at comparable pressures~\cite{ceperley_groundstate_1987}. Exchange between localized sites is a tunneling process suppressed by the small overlap between permutation sectors~\cite{ceperley_exchangefrequencies_1987,ceperley_pathintegrals_1995}, so we expect its contribution to the static structural observables and relative enthalpies considered here to be small. We also note that while nuclear spin statistics (para- vs ortho-hydrogen) are energetically crucial for the rotationally disordered Phase I~\cite{Silvera:1980zz,silvera_validity_1998}, they become secondary in the broken symmetry phases focused on here, where hindered molecular rotations diminish the impact of specific spin configurations. Thus, the spin statistics are also neglected.

To rigorously enforce translational invariance under PBCs, all network inputs are constructed from periodic descriptors.
We encode particle relative positions using sine and cosine of fractional coordinates~\cite{Cassella:2022boh}, and construct single- and two-particle features $\mathbf{h}_{\alpha\beta}^{1}$, where $\alpha,\beta$ denote any pair of particles (electron-electron, nucleus-nucleus, or electron-nucleus). The superscript denotes that it is the input feature of the first layer of the neural network.

The explicit, flexible parametrization of the nuclear wave function $\chi(\mathbf{R})$ is a central feature of our ansatz. The nuclear log-amplitude $\ln \chi(\mathbf{R})$ is constructed from ion-ion features $\mathbf{h}_{I,J}^{1}$ processed by a deep residual network.
We employ a ReZero scheme~\cite{bachlechner2020rezeroneedfastconvergence}, which introduces a trainable scaling parameter initialized to zero. This design allows the model to start from a physically meaningful periodic distance metric and progressively learn many-body corrections $d_{IJ}^{\text{env}}$, thereby capturing anharmonicity in the nuclear wave function. The final log-probability is a sum of squared deviations from learned equilibrium distances $b_{IJ}$ with weight $\sigma_{IJ}$ as follows:
\begin{equation} \label{eq:log_phi}
  \ln \chi(\mathbf{R}) = -\sum_{IJ} \sigma_{IJ}^2 ( d_{IJ}^{\text{env}} - b_{IJ} )^2
\end{equation}.

The electronic wave function is developed based on DeepSolid~\cite{li_initio_2022}, where antisymmetry is enforced through the product of spin-resolved Slater determinants,
\begin{equation}
  \varphi(\mathbf{r},\mathbf{R}) \;=\; \det[\phi^\uparrow]\;\det[\phi^\downarrow].
\end{equation}.
Single-particle orbitals $\phi_{i}^{\alpha}(\mathbf{r}_j;\mathbf{r}_{\neq j})$ are constructed from residual networks acting on electron-ion features $\mathbf{h}_{i,I}^{1}$ and electron-electron features $\mathbf{h}_{i,j}^{1}$, mixing single- and two-body embeddings across layers to encode higher-order correlations.

Although our implementation follows the overall architecture of DeepSolid, we introduce two key modifications to the original framework to enhance training stability. First, we utilize the sine and cosine periodic embedding~\cite{Cassella:2022boh}, discussed previously, replacing the Jastrow factor-based embedding proposed in Ref.~\cite{whiteheadJastrowCorrelationFactor2016} to ensure smoother feature representation. Second, we simplify the electron-features construction process. While the original DeepSolid constructs single-electron features based on the primitive cell, we found this approach led to training instability in hydrogen systems. We, therefore, bypass the primitive cell construction entirely and operate directly within the simulation cell.

\subsection{Constant-pressure optimization}
In the NPT ensemble, the loss function is the enthalpy:
\begin{equation}
  G \;=\; \langle \Psi | \hat{G} | \Psi \rangle
  \;=\; \int \! d\mathbf{r} \, d\mathbf{R} \; |\Psi(\mathbf{r}, \mathbf{R})|^2 \, G_{\text{loc}}(\mathbf{r}, \mathbf{R}),
\end{equation}
where $\hat{G}=\hat{H}+P_{\text{ext}}\Omega$, $P_{\text{ext}}$ is the external pressure, and $\Omega$ is the cell volume; $G_{\text{loc}}(\mathbf{r}, \mathbf{R})=\Psi^{-1}(\mathbf{r}, \mathbf{R})\hat{G}\Psi(\mathbf{r}, \mathbf{R})$ is the local enthalpy.
As shown in Ref.~\cite{Linteau:2024gpe}, the structure of the wave function allows the enthalpy gradient,
\begin{equation}
  \nabla_\Theta G = 2 \,\mathbb{E}\!\left[\left(G_{\text{loc}}-\mathbb{E}[G_{\text{loc}}]\right)
    \nabla_\Theta \ln |\Psi_\Theta|\right] + \mathbb{E}[\nabla_\Theta G_{\text{loc}}],
  \label{eq:gradient_estimator}
\end{equation}
to be decomposed into two distinct contributions. Here, $\Theta$ collectively denotes the neural network parameters $\theta$ and the lattice parameters $L$. When optimizing $\theta$ at fixed $L$, only the first term contributes; conversely, when optimizing $L$ at fixed $\theta$, only the second term contributes.

However, direct evaluation of the second term, $\mathbb{E}[\nabla_\Theta G_{\text{loc}}]$, is computationally expensive. In practice, we avoid computing this costly term explicitly and instead optimize $L$ via a stochastic simulated annealing procedure, as described below.

At each step, a trial lattice deformation \(L \to L'\) is proposed, and the corresponding enthalpy $G'$ is then evaluated. The move is accepted with the Metropolis acceptance probability $\min(1, e^{-\Delta G/\tau})$, where $\Delta G = G' - G$ and $\tau$ is the fictitious temperature, which is used to avoid the risk of trapping in shallow local minima. The fictitious temperature is decreased according to the Lundy-Mees schedule~\cite{lundy_convergence_1986}, $\tau(t) = \frac{\tau_0}{1 + \beta t}$, ensuring ergodicity at early stages and convergence toward low-enthalpy configurations as the annealing progresses. Here, $\tau_0$ is the initial temperature and $\beta$ is the cooling rate.

A subtle point is that the enthalpies $G$ and $G'$ computed in VMC contain intrinsic statistical noise. Using these noisy estimates directly in the Metropolis violates detailed balance. Although the penalty methods can be used to restore exact Boltzmann distribution in the presence of noise~\cite{ceperleyPenaltyMethodRandom1999}, our goal is not to generate an NPT equilibrium distribution but to locate the minimum-enthalpy structure without evaluating high-order derivatives. In this optimization context, strict detailed balance is unnecessary, and the noise can even be beneficial (see Fig.~S1 and the related discussion in the Supplemental Material (SM)~\cite{supplemental}).

\subsection{Sampling}
To evaluate the loss function and its gradient with respect to the neural network parameters $\theta$, we sample electron and nuclear configurations from $p(\mathbf{r},\mathbf{R})=|\Psi(\mathbf{r},\mathbf{R})|^2$. A naive one-shot Metropolis update in the joint electron-nuclear space is inefficient and often exhibits low acceptance rates for nuclear moves due to the vast disparity in energy scales between electrons and nuclei. We, therefore, employ a \emph{Gibbs block sampler}: in each sweep, we (i) sample electronic coordinates $\mathbf{r}$ from $p(\mathbf{r}|\mathbf{R})$ with fixed nuclei, followed by (ii) sample nuclear coordinates $\mathbf{R}$ from $p(\mathbf{R}|\mathbf{r})$ with fixed electrons.
Each conditional update uses Metropolis-Hastings algorithm, with adaptive step sizes to maintain target acceptance.

\section{Results}
\label{sec:result}

We first present benchmark calculations of hydrogen with dynamic protons and then move on to the candidate $Cmcm$ structure for the broken symmetry phase of hydrogen. We will analyze its experimental signatures, as well as its stability in more conventional DFT calculations with the BO approximation. These analyses place our computational discovery in context and highlight the importance of a fully quantum treatment of both electrons and nuclei.

\subsection{Benchmarks}
VMC/DMC has long been a cornerstone method for investigating the phase diagram of dense hydrogen~\cite{ceperley_groundstate_1987, natoli_crystalstructure_1993,natoli_crystalstructure_1995,wang_magneticstructure_1990,holzmannBackflowCorrelationsElectron2003,attaccaliteStableLiquidHydrogen2008,pierleoniTrialWaveFunctions2008,azadi_quantummonte_2013,holzmann_coupled_}. Given the extensive body of literature available for validation, we first benchmark our method against established VMC/DMC results. Calculations were performed for systems of 54 hydrogen atoms arranged in a BCC lattice with fixed volume at $r_s = 1.31$.

Reference data for Slater-Jastrow wave functions using LDA-DFT orbitals (SJ-LDA) and backflow plane-wave orbitals (BF-PW) were taken from Ref.~\cite{holzmann_coupled_}, while neural quantum state (NQS) energies were obtained from Ref.~\cite{linteau_universal_2025}. A detailed comparison is presented in Table~\ref{tab:groundstate}.

Our energies are consistently lower than those of all reference methods, including both traditional and neural quantum state approaches, while maintaining comparable energy variances. This indicates that our ansatz offers enhanced expressivity for the solid hydrogen ground state.

\begin{table}[htbp]
  \caption{Ground-state energies for  $N=54$ hydrogen atoms arranged into a BCC lattice with $r_s = 1.31$ treated in full quantum manner. The digits in the brackets represent the statistical uncertainty of the energy and corresponding variance. Details are provided in Sec.~IV of the SM~\cite{supplemental}.}
  \centering
  \begin{tabular}{@{}ccc@{}}
    \toprule
    Variational ansatz                         & $E/N$ (Ha/atom)       & $\sigma^2/N$ (Ha$^2$/atom) \\
    \midrule
    SJ-LDA (VMC)  \cite{holzmann_coupled_}     & -0.5195(2)            &                            \\
    SJ-LDA (DMC)  \cite{holzmann_coupled_}     & -0.52415(5)           &                            \\
    BF-PW (VMC)    \cite{holzmann_coupled_}    & -0.52194(5)           & 0.025(1)                   \\
    BF-PW (DMC)    \cite{holzmann_coupled_}    & -0.52610(7)           &                            \\
    NQS (VMC)    \cite{linteau_universal_2025} & -0.52854(9)           & \textbf{0.0088(1)}         \\
    \textbf{Present work}                      & \textbf{-0.529674(5)} & 0.01575(5)                 \\
    \bottomrule
  \end{tabular}
  \label{tab:groundstate}
\end{table}

\subsection{The $Cmcm$ candidate structure}
\label{sec:cmcm_structure}
Having validated our framework, we applied it to investigate the structure of solid hydrogen at 130~GPa—corresponding to the experimental pressure range of broken symmetry phase. NPT simulations were performed on a system comprising 64 protons and 64 electrons. We began by optimizing the most promising candidate, $P2_1/c$~\cite{li_highpressurestructures_2024}, whose optimization was initiated at an external pressure of $P_\text{ext}=10$ GPa, and then gradually increased the pressure to 130~GPa. We found that the structure converged to an orthorhombic structure with $Cmcm$ space-group symmetry, both at $P_\text{ext}=95$ GPa and $P_\text{ext}=130$ GPa.

To assess the robustness of this result, we performed independent optimizations starting from several structurally distinct configurations, including $Pca2_1$, $P6_3/m$, and $P6_3/mmc$ (with aligned molecules), as well as the identified $Cmcm$ structure. These calculations yielded two consistent outcomes among the tested initial structures. First, among the tested initial structures, relaxations starting from both $Cmcm$ and $Pca2_1$ converged to the lowest enthalpies, which were significantly lower—by at least 4~mHa per atom—than those of the $P6_3/mmc$ and $P6_3/m$ phases at 130~GPa. Second, the optimization initialized from $Pca2_1$ not only matched the final enthalpy of the $Cmcm$ structure but also evolved into the $Cmcm$ symmetry, differing only slightly in lattice aspect ratios. Similarly, calculations initialized directly in $Cmcm$ remained stable. Finally, a separate optimization with 96 atoms starting from the $Cmcm$ structure was carried to examine the stability of this structure with respect to system size, and the $Cmcm$ lattice remains stable upon scaling. The convergence from multiple unrelated initial configurations suggests that the orthorhombic $Cmcm$ lattice appears to be the enthalpy minimum within our calculation framework. To extract the equilibrium structure from the optimized configurations, we align the first atom in each batch after training.

\begin{figure}
  \centering
  \includegraphics[width=0.8\linewidth]{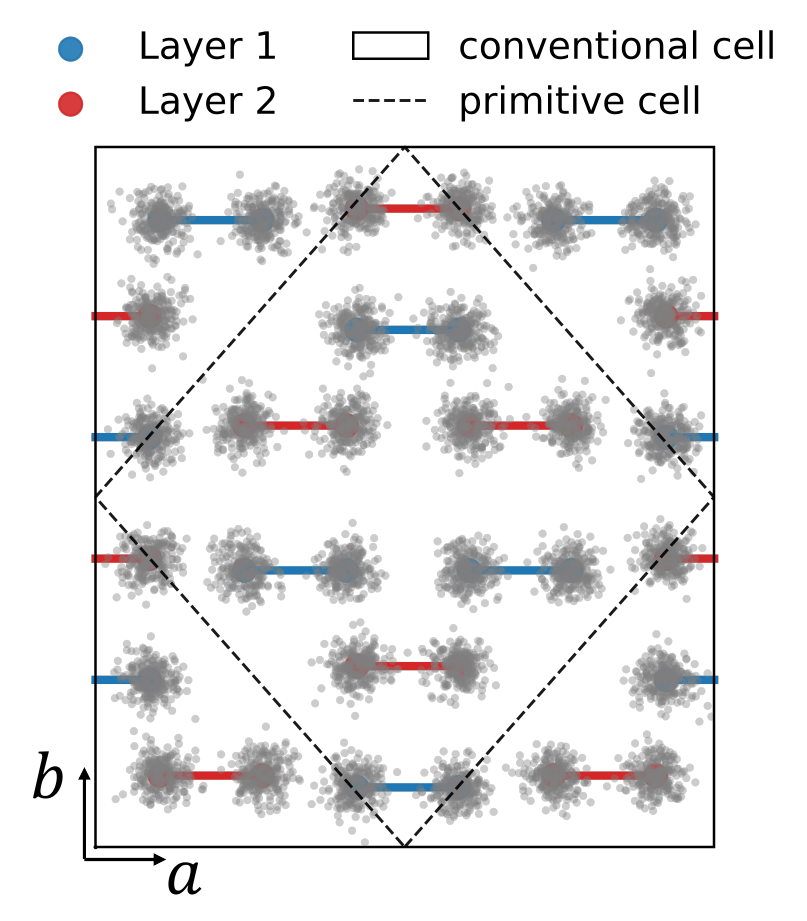}
  \caption{
    The structure with the lowest enthalpy among those tested at 130~GPa for the broken symmetry phase of solid hydrogen. The figure shows the projection of $\mathrm{H}_2$ molecules onto the $ab$ plane. Blue and red lines represent hydrogen molecules in the first and second layers along the $c$ axis, respectively. The black dashed line outlines the primitive cell, while the black solid line indicates the conventional cell. 32 hydrogen atoms in the conventional cell fully occupy four $8g$ Wyckoff positions of the space group $Cmcm$.
  }
  \label{fig:structure}
\end{figure}

Structurally, the discovered orthorhombic $Cmcm$ phase contains 8 $\mathrm{H}_2$ molecules per primitive unit cell (16 molecules in the conventional cell), as illustrated in Fig.~\ref{fig:structure}. The 32 hydrogen atoms fully occupy the four $8g$ Wyckoff positions. Notably, this proposed structure can be viewed as a distortion of the hcp structure found in Phase~I, consistent with the indirect reconstructive mechanism discussed in previous works~\cite{toledano_symmetrybreaking_2009,akahama_evidencexray_2010}. Furthermore, the predicted structural parameters are consistent with experimental measurements. At 130~GPa, our calculated equilibrium volume is 2.063~\AA$^3$/atom, within 4.3\% of the experimental value (2.156~\AA$^3$/atom) reported by~\citet{ji_ultrahighpressureisostructural_2019}. This comparison provides a lattice-metric compatibility check for the $Cmcm$ candidate.

\begin{figure}
  \centering
  \includegraphics[width=\linewidth]{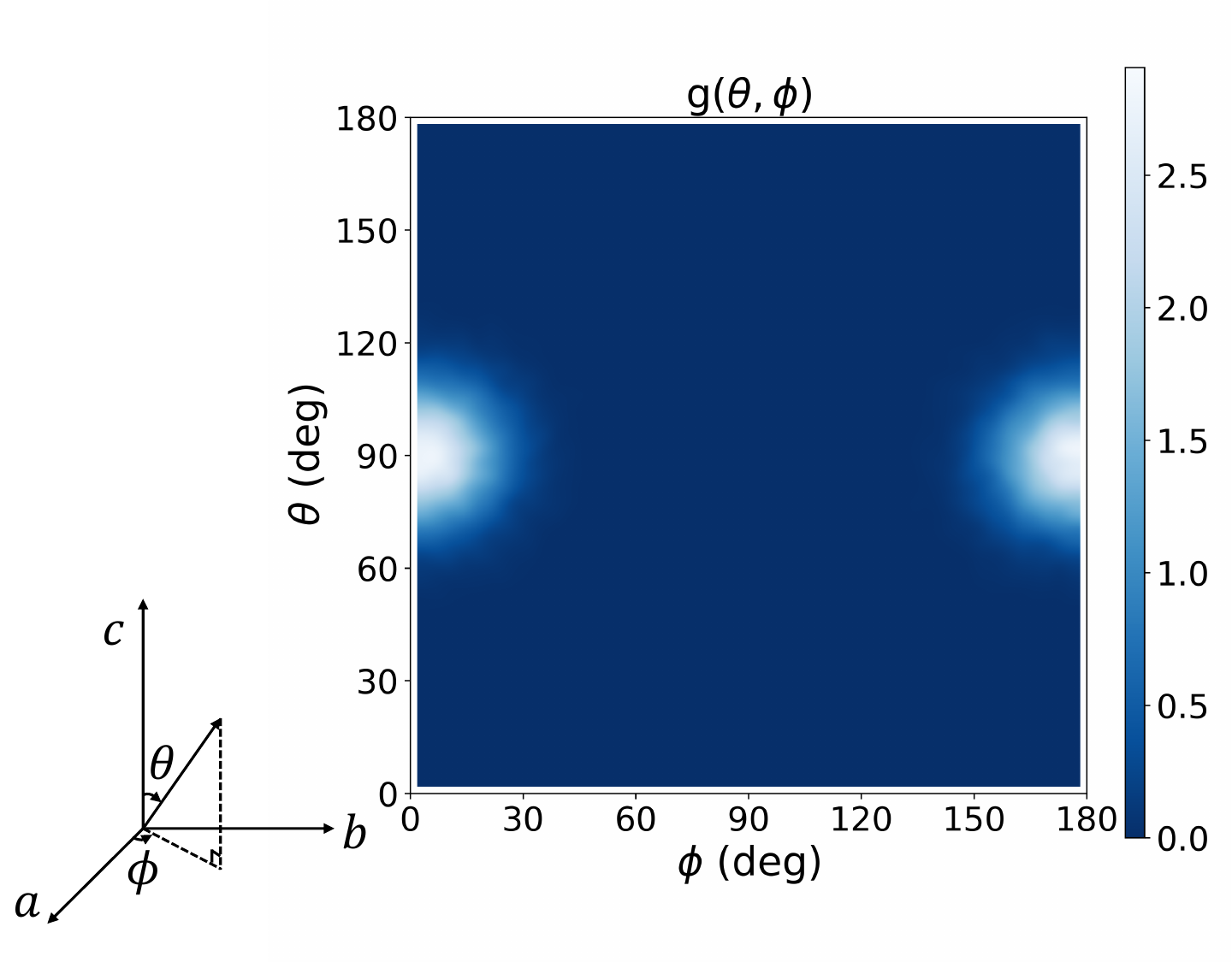}
  \caption{Orientation order of the structure shown in Fig.~\ref{fig:structure}. The angular distribution \(g(\theta,\phi)\) illustrates that preferred orientations of $\mathrm{H}_2$ molecules are within the $ab$ plane. The color bar denotes the normalized density.
  }
  \label{fig:fig2}
\end{figure}

Beyond the lattice symmetry, a defining feature of the identified $Cmcm$ structure is the orientational order of $\mathrm{H}_2$ molecules. Unlike in Phase~I, where the molecules undergo nearly free rotation, the broken symmetry phase exhibits distinct orientational ordering that breaks the high symmetry of the hcp lattice. To quantify this ordering, we computed the joint angular distribution $g(\theta, \phi)$ by accumulating statistics across the sampled configurations. For each molecule, the orientation is characterized by the polar angle $\theta$ and the azimuthal angle $\phi$ as shown in Fig.~\ref{fig:fig2}. The resulting histogram was normalized to yield the probability density. As shown in Fig.~\ref{fig:fig2}, the molecules adopt a layered configuration, lying approximately within the $ab$ plane. This orientational order naturally explains why the observed $c/a=1.586$ in the room-temperature experiment~\cite{ji_ultrahighpressureisostructural_2019} is slightly smaller than the ideal close-packed value of 1.63. A low-temperature measurement at 100~K also reports a similarly reduced axial ratio~\cite{akahama_evidencexray_2010}. Further analysis shows that rotational motion is largely suppressed, while librations persist within the $ac$ plane, confined to an angular range of approximately $60^{\circ} < \theta < 120^{\circ}$.

To obtain a gauge-invariant measure of electron-nuclear coupling, we evaluate the nuclear kinetic energy in the exact-factorization (EF) gauge~\cite{abedi_exact_2010,hunter_conditionalprobability_1975}. Normalizing the conditional electronic state at each nuclear configuration gives $K_{\rm non\mbox{-}BO}^{\rm EF}/K_{\rm BO}^{\rm EF}\approx3.7\%$. This ratio measures the nuclear-coordinate dependence of the normalized conditional electronic state within the optimized full wave function.
\subsection{Experimental signatures}

\subsubsection{X-ray diffraction}

\begin{figure}
  \centering
  \includegraphics[width=\linewidth]{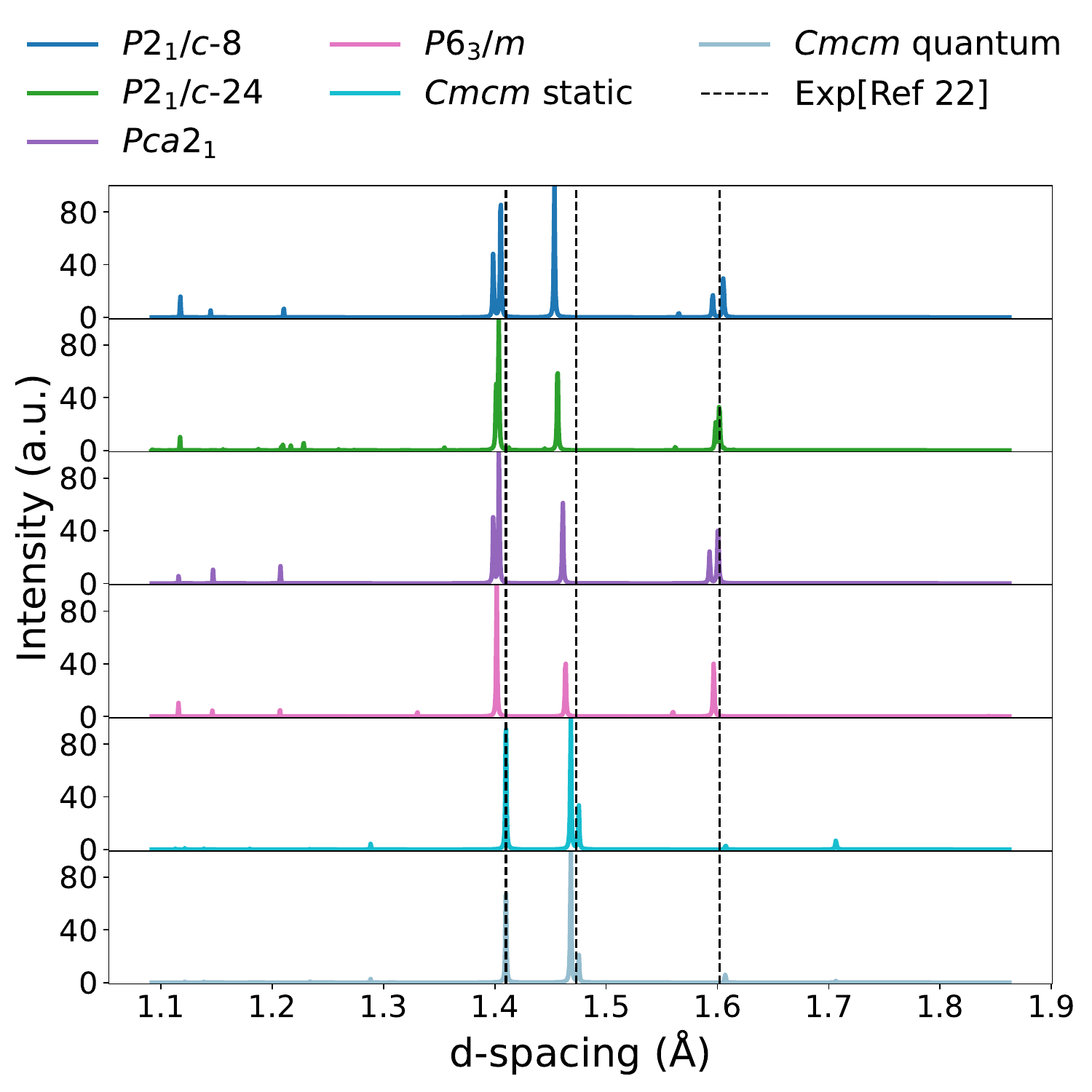}
  \caption{The simulated XRD patterns for various candidate structures at 130~GPa, uniformly rescaled to the experimental volume, compared with the room-temperature observation of Ref.~\cite{ji_ultrahighpressureisostructural_2019}; the NNVMC calculation targets $T=0$. For $Cmcm$, both the static profile and a coherent quantum-density-corrected profile obtained from $|\Psi|^2$ nuclear samples are shown. Details are provided in the SM~\cite{supplemental}. To our knowledge, original XRD profiles (beyond reported peak positions) are unavailable in related literature, with the exception of Ref.~\cite{ji_ultrahighpressurecrystallographic_2025}, which focuses on pressures above 200~GPa. This comparison highlights a gap that calls for further experimental validation.
  }
  \label{fig:xrd}
\end{figure}

We compare the simulated XRD peak positions of the candidate structures with experiment.
As shown in Fig.~\ref{fig:xrd}, after the same uniform volume rescaling, $Cmcm$ gives the closest peak-position match among the tested candidates: the forward and reverse RMS errors are 0.1248 and 0.0036~\AA{} for all selected peaks and 0.0039 and 0.0036~\AA{} within the 1.35--1.65~\AA{} window.
For $Cmcm$, Fig.~\ref{fig:xrd} shows both the static and coherent quantum-density-corrected profiles.
Although the third peak is relatively weak, internal-coordinate refinement and the coherent-density correction described in the SM~\cite{supplemental} slightly enhanced the intensities.
Notably, simulated patterns for other proposed candidate structures, such as $P2_1/c\text{-}8$, $P2_1/c\text{-}24$, $Pca2_1$, and $P6_3/m$, exhibit subtle discrepancies compared to the experimental results. In all cases, the volume is scaled to the experimental volume at the same pressure.
Current experimental precision~\cite{ji_ultrahighpressureisostructural_2019, akahama_evidencexray_2010} is not yet sufficient to unambiguously distinguish several theoretical structures. Nevertheless, together with the NNVMC structural search, this peak-position agreement supports the $Cmcm$ structure.

\subsubsection{Vibrational spectroscopy}

Vibrational spectroscopy provides a symmetry-counting compatibility check against experiment.
Experimentally, the transition from Phase~I to Phase~II is marked by the appearance of new, broadened vibron features in both Raman and IR spectra, absent in the high-symmetry Phase~I, indicating a lowering of lattice symmetry.

We performed a group-theoretical analysis of the vibrational modes for the $Cmcm$ structure to evaluate its spectroscopic characteristics as reference. As summarized in Table~\ref{tab:vibron_phonon}, this symmetry admits four IR-active and four Raman-active vibrons. Although the theoretical mode count exceeds current experimental observations, the $Cmcm$ symmetry remains compatible with the data, as the experimentally observed features can be interpreted as a subset of the theoretically allowed active modes. This interpretation follows the symmetry-counting comparison previously used for candidate structures of solid deuterium~\cite{cui_excitationsorder_1995}. Because the present all-particle NNVMC method does not directly compute Raman or IR frequencies or intensities, the comparison is restricted to symmetry-allowed mode counting.

\begin{table*}[ht]
  \centering
  \caption{Summary of theoretical predictions and experimental observations on IR and Raman modes. The comparison is based on symmetry-allowed mode counting and the observed features may form a subset of the allowed modes. Question marks (?) denote that the corresponding phonon modes are currently unreported or unresolvable. To our knowledge, the IR phonon is not reported in the literature, and the Raman phonon is reported as a single peak in \cite{hemley_lowfrequency_1990}, while \citet{hemley_spectroscopicstudies_1998} observed that this phonon mode becomes extremely weak.
  }
  \label{tab:vibron_phonon}
  \begin{tabular}{ccccc}
    \toprule
    Space group                                                                              & IR vibron        & Raman vibron      & IR phonon                     & Raman phonon                         \\
    \midrule
    $Pca2_1$ \cite{cui_excitationsorder_1995}                                                & $A_1+B_1+B_2$    & $A_1+A_2+B_1+B_2$ & $2A_1+2B_1+2B_2$              & $2A_1+3A_2+2B_1+2B_2$                \\
    $P2_1/c$   \cite{cui_excitationsorder_1995}                                              & $A_u+B_u$        & $A_g+B_g$         & $2A_u+B_u$                    & $3A_g+3B_g$                          \\
    $P6_3/m$    \cite{cui_excitationsorder_1995}                                             & $E_{1u}$         & $2A_g+E_{2g}$     & $A_u+2E_{1u}$                 & $2A_g+E_{1g}+3E_{2g}$                \\
    $Cmcm$                                                                                   & $3B_{2u}+B_{3u}$ & $3A_{g}+B_{1g}$   & $3B_{2u} + 6B_{1u} + 3B_{3u}$ & $4B_{1g} + 7B_{3g} + 4A_g + 5B_{2g}$ \\
    \midrule
    Experiments  \cite{Mao:1994zz,hemley_spectroscopicstudies_1998,hemley_lowfrequency_1990} & $2$              & $1$               & ?                             & ?                                    \\
    \bottomrule
  \end{tabular}
\end{table*}
The analysis of the lattice phonon modes presents a greater challenge due to limited experimental data in the low-frequency region. As shown in Table~\ref{tab:vibron_phonon}, the $Cmcm$ structure allows for a rich spectrum of phonon modes. However, no IR phonon modes have been reported in the literature to our knowledge. Similarly, the Raman phonon features are elusive; while earlier work reported a single peak~\cite{hemley_lowfrequency_1990}, subsequent studies observed that this phonon mode becomes extremely weak~\cite{hemley_spectroscopicstudies_1998}. The absence of these theoretical peaks in current data may be attributable to weak signal intensities~\cite{zhang_brokensymmetry_2007} or peak overlapping characteristic of high-pressure hydrogen experiments, rather than an exclusion of the $Cmcm$ symmetry.

\subsection{Stability analysis with static DFT calculations}
\label{subsec:dft_stability}

To understand why the $Cmcm$ structure identified in our calculation was not found in previous DFT-based searches, we performed static DFT calculations using the PBE functional. We conducted variable-cell relaxations (\texttt{vc-relax}) on the $Cmcm$ structure alongside other competitive candidates ($P2_1/c\text{-}8$, $Pca2_1$, $P6_3/m$ with unit cells containing 64 atoms, and $P2_1/c\text{-}24$ with 48 atoms). Convergence was strictly verified with Monkhorst-Pack $\mathbf{k}$-point meshes up to $8\times8\times8$, ensuring that finite-size errors do not affect our conclusions.

Our static DFT results reveal a significant discrepancy with the NNVMC $Cmcm$ result. Across all tested structures, the relaxed space groups remained unchanged during the \texttt{vc-relax}. Apart from $Cmcm$, all other candidates exhibit very similar final enthalpies per atom at a given $\mathbf{k}$-point mesh. In contrast to our calculation, the static enthalpy of $Cmcm$ is not the global minimum, it lies approximately $1.1$~mHa/atom higher than $Pca2_1$. To further assess the dynamical stability, we computed the phonon spectrum. While other candidates are dynamically stable, the relaxed $Cmcm$ structure exhibits negative (imaginary) frequencies at the $\Gamma$ point. This shows that, on the classical BO surface, the $Cmcm$ lattice is a saddle point rather than a stable minimum.

The fact that $Cmcm$ is a saddle point with higher static enthalpy explains its absence in conventional DFT searches. There are several possibilities for such discrepancy as the static DFT calculation involves multiple approximations that we have relaxed in our many-body calculation. First, there are DFT functional errors~\cite{azadi_fatedensity_2013}. Second, it is well known that strong NQEs, including ZPE, lattice anharmonicity, and even beyond BO effects, may overturn classical structural predictions. A detailed assessment is beyond the scope of the present paper. In the following, we speculate on the effects of lattice anharmonicities and beyond BO contributions in stabilizing the discovered $Cmcm$ phase.

First, NQEs such as ZPE and anharmonicity play a critical role. Although the static enthalpy of $Cmcm$ is higher than that of $Pca2_1$, the difference is small ($\sim 1.1$~mHa/atom). Once NQEs are considered, the energetic ordering can be reversed. The ZPE of solid hydrogen at this pressure is estimated to be around $10$~mHa/atom based on harmonic phonon calculations~\cite{gorai_abinitio_2023}. Therefore, even a moderate estimate that different competing structures differ by $10\%$ in their ZPE is comparable to the $\sim 1$~mHa/atom  difference between the static-lattice enthalpy of $Cmcm$ and $Pca2_1$ structures.

Second, our framework explicitly captures non-BO effects. As detailed in Sec.~\ref{sec:cmcm_structure}, the EF-gauge non-BO contribution to the kinetic energy in the $Cmcm$ phase is approximately 0.22~mHa/atom. Although smaller than the ZPE scale, this term may also contribute to the relative structural stability.

Taken together, these results demonstrate that $Cmcm$ is not stable at the static-lattice level within the PBE functional. Instead, its emergence as the enthalpy-minimizing structure in our framework is driven by strong NQEs, including anharmonicity and non-BO effects. This physical picture is consistent with a large body of recent work on hydrogen-rich systems. \citet{errea_quantum_2020} shows that quantum fluctuations reshape the BO landscape of LaH$_{10}$, collapsing multiple classical minima into a single high-symmetry phase; and~\citet{monserrat_structure_2018} demonstrates how anharmonic and quantum effects stabilize otherwise classically unstable high-pressure hydrogen phases. Similarly, our findings also suggest that strong protonic quantum motion may effectively lift the saddle point on the BO surface, transforming it into a minimum on the full quantum energy landscape.

These observations not only explain why previous DFT-based searches missed the $Cmcm$ structure but also highlight the importance of a fully anharmonic and beyond BO description, such as that provided by our method, for correctly determining the phase diagram of dense solid hydrogen.

\section{Discussion and Conclusion}
\label{sec:conclusion}
The present work revisits one of the longest-standing problems in high-pressure physics—the structural nature of the broken symmetry phase of solid hydrogen. By employing NNVMC framework, which treats electrons and nuclei on an equal footing and directly minimizes enthalpy in the NPT ensemble, we identify an orthorhombic molecular lattice with $Cmcm$ symmetry as the lowest-enthalpy structure in our NNVMC search at 130~GPa. Agreement with the equation of state, the closest XRD peak-position match, and Raman and infrared symmetry compatibility provide three complementary consistency checks for this candidate. Taken together, these results indicate that $Cmcm$ is a potentially competitive candidate for the broken symmetry phase and motivate targeted experimental tests.

An intriguing question is whether the structural transition of the $\mathrm{H}_2$ hcp structure coincides with the orientational ordering $\mathrm{H}_2$ molecules. Our analysis suggests that these may be two distinct symmetry breaking transitions. As noted in Ref.~\cite{toledano_symmetrybreaking_2009}, the $Cmcm$ symmetry is a maximal subgroup of the Phase~I $P6_3/mmc$ space group. In the model proposed by Toledano \textit{et al.}, H atoms in Phase~I partially occupy the $4f$ and $12k$ sites, reflecting the orientational disorder of freely rotating molecules. They proposed that compression induces a transition to a partially ordered $Cmcm$ phase, where fractional occupancy persists (e.g., half-filled $8f$ and $16h$ sites), leading to orientational degeneracy and potential topological frustration that could drive an incommensurate modulation.

In contrast, our calculations suggest that the broken symmetry phase is fully ordered, characterized by full occupancy of the $8g$ Wyckoff positions of the $Cmcm$ space group. The subgroup pathway inferred from our calculations differs from that of Ref.~\cite{toledano_symmetrybreaking_2009}. To link our $Cmcm$ structure to Phase~I, we consider a restricted $P6_3/mmc$ model where molecules are confined to the $ab$ plane (corresponding to partially occupied $6h$ sites, derived from $12k$ with $z=1/4$ and the $4f$ sites are vacated). Under the group-subgroup reduction $P6_3/mmc \rightarrow Cmcm$, these partially occupied $6h$ sites split into the fully occupied $8g$ positions of the $Cmcm$ structure. Consequently, the resulting phase is free of orientational degeneracy, eliminating the mechanism previously hypothesized to drive incommensurate distortion. Furthermore, unlike many previously proposed candidate structures where molecular centers of mass (COM) retain the hcp lattice positions of Phase~I, the COMs in our discovered $Cmcm$ structure exhibit a more complex structure. While the COMs no longer form a simple hcp lattice, the structure can be fundamentally viewed as a distortion of hcp. This symmetry analysis hints at a possible two-step transformation sequence involving an intermediate state that retains $P6_3/mmc$ lattice symmetry but lacks free rotation. Further investigation into the coupling between lattice symmetry breaking and orientational ordering is warranted.

On the methodological front, similar to the beyond BO neural wave function approach in Ref.~\cite{linteau_universal_2025}, we have developed a general first principle framework for predicting ground states of quantum materials beyond BO approximation within NPT ensemble. The necessity of such an approach is underscored by our stability analysis in Sec.~\ref{subsec:dft_stability}, which reveals that the $Cmcm$ structure is a saddle point on the static PES. It is stabilized only by strong nuclear quantum effects—specifically, the interplay of zero-point motion and nonadiabatic coupling. By treating variable cell parameters as variational degrees of freedom and minimizing the enthalpy of a coupled electron-nuclear wave function, our calculations capture these critical stabilizing mechanisms that standard DFT methods inherently miss. Consequently, the optimal crystal structure emerges naturally as an intrinsic property of the many-body quantum ground state. Scalable to systems with hundreds of quantum particles, this approach provides a versatile tool for investigating systems where quantum fluctuations reshape the energy landscape, making it particularly well suited for exploring higher-pressure hydrogen phases, high-$T_c$ hydrides, and other light-element materials under extreme conditions.

Localization and prior molecular QMC results indicate that the proton-exchange correction neglected here is small for the static observables considered. The present calculation targets the low-temperature BSP/Phase~II regime. Extending direct NNVMC comparisons to A-phase/Phase~III structures is an important direction for future work.

In conclusion, our full quantum many-body calculations identify orthorhombic $Cmcm$ as a potentially competitive candidate for the broken symmetry phase within the structures examined. This finding sheds new light on a long-standing problem in high-pressure physics, showing consistency with available structural, spectroscopic, and thermodynamic observations, and motivates further theoretical and experimental assessment. Beyond identifying this candidate, the present work explores a general computational framework for predicting quantum materials under finite pressure, treating electronic and nuclear degrees of freedom on an equal footing. Crucially, our results show that a full nonadiabatic quantum treatment, capturing anharmonicity and beyond BO effects, plays an important role in describing the phase diagram of dense hydrogen. These results motivate further experimental tests of $Cmcm$ in the broken symmetry phase.

\begin{acknowledgments}
  We thank David Ceperley, Yuichi Akahama, Hao Xie, Mao Su, Gengyuan Hu, Zi-Xiao Zhang, Cheng Chen, Hongfang Liu, Jingxu Chen for helpful discussions. This work is supported by the National Natural Science Foundation of China under Grants No. T2225018, No. 92270107,  No. 12188101, No. T2121001, No. 12504289, the Strategic Priority Research Program of the Chinese Academy of Sciences under Grants No. XDB0500000, and the National Key Projects for Research and Development of China Grants No. 2021YFA1400400, New Generation Artificial Intelligence-National Science and Technology Major Project (Grants No. 2025ZD0121802), Quantum Science and Technology-National Science and Technology Major Project (Grants No. 2021ZD0302400).

\end{acknowledgments}

\section*{Data Availability}
The source code and optimized candidate structures reported in this work are publicly available in the \textsc{BornFree} repository~\cite{chai_bornfree_repository_2025}.

\bibliography{main}
\clearpage
\onecolumngrid

\begin{center}
  \textbf{\large Supplemental Material for Revisiting the Broken Symmetry Phase of Solid Hydrogen: A Neural Network Variational Monte Carlo Study}
\end{center}

\begin{center}
  \large
  Shengdu Chai\textsuperscript{1,2},
  Chen Lin\textsuperscript{3},
  Xinyang Dong\textsuperscript{4},
  Yuqiang Li\textsuperscript{2}, \\
  Wanli Ouyang\textsuperscript{2,5},
  Lei Wang\textsuperscript{4},
  and X.C. Xie\textsuperscript{1,6,7}
\end{center}

\begin{center}
  \small \itshape
  \textsuperscript{1}Interdisciplinary Center for Theoretical Physics and Information Sciences (ICTPIS), Fudan University, Shanghai 200433, China\\
  \textsuperscript{2}Shanghai Artificial Intelligence Laboratory, Shanghai 200232, China\\
  \textsuperscript{3}Department of Engineering Science, University of Oxford, Oxford OX1 3PJ, United Kingdom\\
  \textsuperscript{4}Beijing National Laboratory for Condensed Matter Physics and Institute of Physics, Chinese Academy of Sciences, Beijing 100190, China\\
  \textsuperscript{5}Department of Information Engineering, The Chinese University of Hong Kong, Hong Kong SAR HKG, China\\
  \textsuperscript{6}International Center for Quantum Materials, School of Physics, Peking University, Beijing 100871, China\\
  \textsuperscript{7}Hefei National Laboratory, Hefei 230088, China
\end{center}

\setcounter{equation}{0}
\setcounter{figure}{0}
\setcounter{table}{0}
\setcounter{section}{0}
\setcounter{alg}{0}

\renewcommand{\theequation}{S\arabic{equation}}
\renewcommand{\theHequation}{S.\arabic{equation}}
\renewcommand{\thefigure}{S\arabic{figure}}
\renewcommand{\theHfigure}{S.\arabic{figure}}
\renewcommand{\thetable}{S\arabic{table}}
\renewcommand{\theHtable}{S.\arabic{table}}
\providecommand{\theHalg}{}
\renewcommand{\theHalg}{S.\arabic{alg}}
\setcounter{secnumdepth}{3}

\section{Neural Network Wavefunction}

To satisfy the PBCs required by the simulation cell, we map Cartesian coordinates to periodic descriptors via fractional coordinates~\cite{Cassella:2022boh}. Let $\mathbf{r}_{ij} = \mathbf{r}_i - \mathbf{r}_j$ be a relative displacement vector. We decompose this vector into the lattice basis $\{\mathbf{a}_k\}$ as $\mathbf{r}_{ij} = \sum_{k=1}^3 \omega_{ij,k}\mathbf{a}_k$. The periodic distance is computed using the lattice metric tensor $S_{kl}=\mathbf{a}_{k}\!\cdot\!\mathbf{a}_{l}$ as:
\begin{equation}
  d(\mathbf{r}) \;=\;
  \Big[(1-\cos\omega_{k})S_{kl}(1-\cos\omega_{l})+\sin\omega_{k}\,S_{kl}\sin\omega_{l}\Big]^{1/2}.
  \label{eq:periodic_dist}
\end{equation}
The input feature vectors $\mathbf{h}_{\alpha\beta}^{1}$ for the first neural network layer are constructed by concatenating the periodic embedding of relative positions with this distance metric:
\begin{equation}
  \mathbf{h}_{\alpha\beta}^{1} \;=\;
  \Big\{ \sum_{k}\sin(\omega_{\alpha\beta}^{k})\mathbf{a}_{k}, \;
  \sum_{k}\cos(\omega_{\alpha\beta}^{k})\mathbf{a}_{k}, \;
  d(\omega_{\alpha\beta}) \Big\},
  \label{eq:features}
\end{equation}
where $\alpha,\beta$ denote any pair of particles (electron--electron, nucleus--nucleus, or electron--nucleus).

The pseudocode for the nuclear wavefunction and electron wavefunction is given in \autoref{alg:nuclearwavefunction} and \autoref{alg:elecwavefunction}, respectively. As for the sampling algorithm, the two-stage Gibbs sampler is summarized in \autoref{alg:gibbs}, with the original one-shot Metropolis scheme provided for reference in \autoref{alg:mcmc}
\begin{figure}[htbp]
  \begin{ruledalgorithm}
    \algcaption{Pseudocode of $\ln \chi(\mathbf{R})$ for nuclear wavefunction}
    \vspace{2pt}\hrule height 0.5pt \vspace{6pt}
    \begin{algorithmic}[1]
      \State \textbf{Input:} nuclear positions $\{ \mathbf{R}_I \}$
      \State \textbf{Input:} lattice vector \(\{ \mathbf{a_1}, \mathbf{a_2}, \mathbf{a_3} \}\) of simulation cell
      \State \textbf{Input:} reciprocal lattice vector \(\{ \mathbf{b_1}, \mathbf{b_2}, \mathbf{b_3} \}\) of simulation cell
      \State \textbf{Output:} log of nuclear wavefunction $\ln \chi(\mathbf{R})$
      \For{each atom $I,J$}
      \State $\omega_{IJ}^{k}=(\mathbf{R}_{I}-\mathbf{R}_{J})\cdot \mathbf{b}_{k}$
      \EndFor
      \For{each atom $I,J$}
      \State $\mathbf{h}_{I,J}^{1} = \{\sum_{k}\sin(\omega_{IJ}^{k})\mathbf{a_{k}}, \sum_{k}\cos(\omega_{IJ}^{k})\mathbf{a_{k}},d(\omega_{IJ})\}$
      \EndFor
      \For{each layer $l$}
      \State $\mathbf{h}^{l+1}_{IJ} = \tanh(\mathbf{\Pi}^l_{\text{atom}} \cdot \mathbf{h}^{l}_{IJ} + \mathbf{c}^l_{\text{atom}})$
      \State $\mathbf{h}^{l+1}_{IJ} = \frac{1}{\sqrt{2}}(\mathbf{h}^{l}_{IJ} + \mathbf{h}^{l+1}_{IJ})$ \Comment{Residual connection}
      \EndFor

      \State $\mathbf{d}^{\text{pred}}_{IJ} = \mathbf{\Pi}^{\text{dist}}_{\text{atom}} \cdot \mathbf{h}^{L}_{IJ} + \mathbf{c}^{\text{dist}}_{\text{atom}}$ \Comment{Shape (natom, natom, 1)}
      \State $d_{IJ}^{\text{env}} = \alpha* \mathbf{d}^{\text{pred}}_{IJ} + d(\omega_{IJ})$ \Comment{Rezero}

      \State $\ln \chi(\mathbf{R}) = -\sum_{IJ} \sigma_{IJ}^2 ( d_{IJ}^{\text{env}} - b_{IJ} )^2$ \Comment{b is param of network}
    \end{algorithmic}
    \label{alg:nuclearwavefunction}
  \end{ruledalgorithm}
\end{figure}
\begin{figure}[htbp]
  \begin{ruledalgorithm}
    \algcaption{Pseudocode of electron wavefunction}
    \vspace{2pt}\hrule height 0.5pt \vspace{6pt}
    \begin{algorithmic}[1]
      \State \textbf{Input:} electron positions $\{ \mathbf{r}_1^\uparrow, \ldots, \mathbf{r}_{n_{\uparrow}}^\uparrow, \mathbf{r}_1^\downarrow, \ldots, \mathbf{r}_{n_{\downarrow}}^\downarrow \}$
      \State \textbf{Input:} nuclear positions \(\{ \mathbf{R}_I \}\) in the simulation cell
      \State \textbf{Input:} lattice vector \(\{ \mathbf{a_1}, \mathbf{a_2}, \mathbf{a_3} \}\) of simulation cell
      \State \textbf{Input:} reciprocal lattice vector \(\{ \mathbf{b_1}, \mathbf{b_2}, \mathbf{b_3} \}\) of simulation cell
      \State \textbf{Input:} occupied \(\{ \mathbf{k}_i \}\)
      \State \textbf{Output:} wavefunction ansatz $\varphi(\mathbf{r},\mathbf{R})$
      \For{each electron $i,j$, atom $I$}
      \State $\omega_{iI}^{k}=(\mathbf{r}_{i}-\mathbf{R}_{I})\cdot \mathbf{b}_{k}$,
      \State $\omega_{ij}^{k}=(\mathbf{r}_{i}-\mathbf{r}_{j})\cdot \mathbf{b}_{k}$
      \EndFor
      \For{each electron $i,j$, atom $I$}
      \State $\mathbf{h}_{i,I}^{1} = \{\sum_{k}\sin(\omega_{iI}^{k})\mathbf{a_{k}}, \sum_{k}\cos(\omega_{iI}^{k})\mathbf{a_{k}},d(\omega_{iI})\}$
      \State $\mathbf{h}_{i,j}^{1} = \{\sum_{k}\sin(\omega_{ij}^{k})\mathbf{a_{k}}, \sum_{k}\cos(\omega_{ij}^{k})\mathbf{a_{k}},d(\omega_{ij})\}$
      \EndFor
      \For{each layer $l$}
      \State$\mathbf{g}^{l,\uparrow}=\frac{1}{n_{\uparrow}}\sum_{i}\mathbf{h}_{i,I}^{l,\uparrow}$
      \State$\mathbf{g}^{l,\downarrow}=\frac{1}{n_{\downarrow}}\sum_{i}\mathbf{h}_{i,I}^{l,\downarrow}$
      \For{each electron $i$, spin $\alpha$}
      \State$\mathbf{g}^{l,\alpha,\uparrow}_{i}=\frac{1}{n_{\uparrow}}\sum_{j}\mathbf{h}_{i,j}^{l,\alpha,\uparrow}$
      \State$\mathbf{g}^{l,\alpha,\downarrow}_{i}=\frac{1}{n_{\downarrow}}\sum_{j}\mathbf{h}_{i,j}^{l,\alpha,\downarrow}$
      \State$\mathbf{f}_{i}^{l,\alpha}$ = concat($\mathbf{h}_{i}^{l,\alpha}, \mathbf{g}^{l,\uparrow}, \mathbf{g}^{l,\downarrow}, \mathbf{g}^{l,\alpha,\uparrow}_{i}, \mathbf{g}^{l,\alpha,\downarrow}_{i}$)
      \State$\mathbf{h}_{i}^{l+1,\alpha} = \tanh(V^l \cdot \mathbf{f}_{i}^{l,\alpha} + \mathbf{b}^{l}) + \mathbf{h}_{i}^{l,\alpha}$
      \State$\mathbf{h}_{i,j}^{l+1,\alpha,\beta}=\tanh(W^l \cdot \mathbf{h}_{i,j}^{l,\alpha,\beta} + \mathbf{c}^{l}$) + $\mathbf{h}_{i,j}^{l,\alpha,\beta}$
      \EndFor
      \EndFor
      \For{each electron $i,j$,spin $\alpha$}
      \State $u_{\mathbf{k_{i}}}^{\alpha}(\mathbf{r_{j}};\mathbf{r_{\neq j}}) =\text{Orb}^{Re}_{j,\alpha}\mathbf{h}_{i,I}^{L}+ i \text{Orb}^{Im}_{j,\alpha}\mathbf{h}_{i,I}^{L}$
      \State $p_{i,j}^{\alpha}$ = $\exp(i \mathbf{k_i} \cdot \mathbf{r_j^\alpha})$
      \State $env_{i,j}^{\alpha}$ = $\sum_{I} \pi_{i,I}^{\alpha} \exp[-\sigma_{i,I}^{\alpha} d(\omega_{i,I})]$
      \State $\phi_{i,j}^{\alpha}$ = $p_{i,j}^{\alpha} u_{\mathbf{k_{i}}}^{\alpha}(\mathbf{r_{j}};\mathbf{r_{\neq j}}) env_{i,j}^{\alpha}$
      \EndFor
      \State $\varphi(\mathbf{r},\mathbf{R}) = \text{Det}[\phi^\uparrow] \text{Det}[\phi^\downarrow]$
    \end{algorithmic}
    \label{alg:elecwavefunction}
  \end{ruledalgorithm}
\end{figure}
\begin{figure}[htbp]
  \begin{ruledalgorithm}
    \algcaption{Gibbs Block Sampling for Joint Electron-Nucleus Configuration}
    \vspace{2pt}\hrule height 0.5pt \vspace{6pt}
    \label{alg:gibbs}
    \begin{algorithmic}[1]
      \State \textbf{Input:} initial electron positions $\mathbf{r}^{(0)}$
      \State \textbf{Input:} initial nuclear positions $\mathbf{R}^{(0)}$
      \State \textbf{Input:} unnormalized probability density $p(\mathbf{r}, \mathbf{R}) = |\Psi(\mathbf{r}, \mathbf{R})|^2$
      \State \textbf{Input:} proposal distributions $q(\mathbf{R} | \mathbf{\hat{R}})$, $q(\mathbf{r} | \mathbf{\hat{r}})$
      \State \textbf{Input:} number of iterations $T$
      \State \textbf{Input:} number of blocks $B$
      \State \textbf{Output:} samples $\{ \mathbf{R}^{(t)},\mathbf{r}^{(t)} \} \sim p(\mathbf{R}, \mathbf{r})$
      \State Initialize $\mathbf{r}_{\text{prev}} \leftarrow \mathbf{r}^{(0)}$, $\mathbf{R}_{\text{prev}} \leftarrow \mathbf{R}^{(0)}$
      \For{$b = 1$ to $B$}
      \For{$\tau = 1$ to $T$}
      \State Propose $\mathbf{R}^* \sim q(\mathbf{R} | \mathbf{R}_{\text{prev}})$
      \State Draw $u \sim \mathcal{U}(0,1)$
      \State Compute acceptance ratio $\alpha = \frac{p(\mathbf{R}^* | \mathbf{r}_{\text{prev}})}{p(\mathbf{R}_{\text{prev}} | \mathbf{r}_{\text{prev}})}$
      \If{$u < \alpha$}
      \State $\mathbf{R}_{\text{prev}} \leftarrow \mathbf{R}^*$
      \EndIf
      \EndFor

      \For{$\tau = 1$ to $T$}
      \State Propose $\mathbf{r}^* \sim q(\mathbf{r} | \mathbf{r}_{\text{prev}})$
      \State Draw $u \sim \mathcal{U}(0,1)$
      \State Compute acceptance ratio $\alpha = \frac{p(\mathbf{r}^* | \mathbf{R}_{\text{prev}})}{p(\mathbf{r}_{\text{prev}} | \mathbf{R}_{\text{prev}})}$
      \If{$u < \alpha$}
      \State $\mathbf{r}_{\text{prev}} \leftarrow \mathbf{r}^*$
      \EndIf
      \EndFor
      \EndFor
      \State Record: $\mathbf{R}^{(t)} \leftarrow \mathbf{R}_{\text{prev}},\ \mathbf{r}^{(t)} \leftarrow \mathbf{r}_{\text{prev}}$
    \end{algorithmic}
  \end{ruledalgorithm}
\end{figure}

\begin{figure}[htbp]
  \begin{ruledalgorithm}
    \algcaption{Metropolis Sampling for Joint Electron-Nucleus Configuration}
    \vspace{2pt}\hrule height 0.5pt \vspace{6pt}
    \label{alg:mcmc}
    \begin{algorithmic}[1]
      \State \textbf{Input:} initial electron positions $\mathbf{r}^{(0)}$
      \State \textbf{Input:} initial nuclear positions $\mathbf{R}^{(0)}$
      \State \textbf{Input:} unnormalized probability density $p(\mathbf{r}, \mathbf{R}) = |\Psi(\mathbf{r}, \mathbf{R})|^2$
      \State \textbf{Input:} proposal distributions $q(\mathbf{R} | \mathbf{\hat{R}})$, $q(\mathbf{r} | \mathbf{\hat{r}})$
      \State \textbf{Input:} number of iterations $T$
      \State \textbf{Output:} samples $\{ \mathbf{R}^{(t)}, \mathbf{r}^{(t)} \} \sim p(\mathbf{r}, \mathbf{R})$
      \State Initialize $\mathbf{r}_{\text{prev}} \leftarrow \mathbf{r}^{(0)}$, $\mathbf{R}_{\text{prev}} \leftarrow \mathbf{R}^{(0)}$
      \For{$\tau = 1$ to $t$}
      \State Propose $\mathbf{r}^* \sim q(\mathbf{r} | \mathbf{r}_{\text{prev}})$
      \State Propose $\mathbf{R}^* \sim q(\mathbf{R} | \mathbf{R}_{\text{prev}})$
      \State Draw $u \sim \mathcal{U}(0,1)$
      \State Compute acceptance ratio $\alpha = \frac{p(\mathbf{r}^*, \mathbf{R}^*)}{p(\mathbf{r}_{\text{prev}}, \mathbf{R}_{\text{prev}})}$
      \If{$u < \alpha$}
      \State $\mathbf{r}_{\text{prev}} \leftarrow \mathbf{r}^*$,\quad $\mathbf{R}_{\text{prev}} \leftarrow \mathbf{R}^*$
      \EndIf
      \EndFor
      \State Record: $\mathbf{r}^{(t)} \leftarrow \mathbf{r}_{\text{prev}},\ \mathbf{R}^{(t)} \leftarrow \mathbf{R}_{\text{prev}}$
    \end{algorithmic}
  \end{ruledalgorithm}
\end{figure}

\section{Enthalpy operator and coordinate map}
Rather than treating lattice parameters as variational degrees of freedom in $\Psi$, we incorporate them into the enthalpy operator and sample in unit-scaled (fractional) coordinates~\cite{Linteau:2024gpe}. We define the set of all real-space particle coordinates as $\mathcal{X} = \{\mathbf{r}_i\} \cup \{\mathbf{R}_I\}$ and the corresponding set of fractional coordinates as $\mathcal{S} = \{\mathbf{s}_i\} \cup \{\mathbf{S}_I\}$, where all $\mathbf{s}_i, \mathbf{S}_I \in [0,1]^3$. The mapping from fractional to real coordinates is given by $\mathcal{X} = L^{\top} \mathcal{S}$, where $L$ is the lattice matrix and $L^{\top}\mathcal{S}=(L^{\top}s_1,\ldots,L^{\top}s_{N},L^{\top}S_1,\ldots,L^{\top}S_{N})$. Defining $g=LL^{\top}$ and $g^{-1}=(L^{-1})^{\top}L^{-1}$, the local estimator becomes,
\begin{equation}
  \begin{aligned}
    G_{\text{loc}}(\mathcal{S}) & = K_{\text{loc}}(\mathcal{S}) + V_{\text{loc}}(\mathcal{S})                                                                                             \\
    K_{\text{loc}}(\mathcal{S}) & =  -\sum_{p}\frac{1}{2m_p}[\sum_{\alpha,\beta}g^{-1}_{\alpha,\beta}\frac{\partial^2}{\partial_{s_p}^{\alpha}\partial_{s_p}^{\beta}}\ln\Psi(\mathcal{S}) \\
                                                & +(L^{-1}\nabla_{s_p}\ln\Psi(\mathcal{S}))^2] \\
    V_{\text{loc}}(\mathcal{S}) & = V(L^{\top}\mathcal{S}) + P \Omega = V(L^{\top}\mathcal{S}) + P \det L
  \end{aligned}
\end{equation}
where $m_p=\begin{cases}
    1, & p\leq N    \\
    M, & N<p\leq 2N
  \end{cases}$. The lattice is characterized by six parameters $(a,b,c,\alpha,\beta,\gamma)$, with $a,b,c$ the lengths of the three basis vectors and $\alpha,\beta,\gamma$ the angles between them. The lattice matrix $L$ is written as
\begin{equation}
  L=\begin{pmatrix}
    a           & 0                       & 0                      \\
    b\cos\gamma & b\sin\gamma             & 0                      \\
    c\cos\beta  & -c\sin\beta\cos\alpha^* & c\sin\beta\sin\alpha^*
  \end{pmatrix}, \qquad
  \cos\alpha^*=\frac{\cos\beta\cos\gamma-\cos\alpha}{\sin\beta\sin\gamma}.
  \label{eq:lattice}
\end{equation}

For clarity, we assume the wavefunction in the NVT ensemble has the form $\Psi_{\theta \cup L_{\text{fixed}}}(\mathcal{X})_{\text{NVT}}$. The NPT wavefunction is defined via the NVT form in the unit cube,
\begin{equation}
  \Psi_{\theta}(\mathcal{S})_{\text{NPT}} \equiv \Psi_{\theta \cup L=I}(\mathcal{S})_{\text{NVT}}.
  \label{eq:wavefunction_npt}
\end{equation}
so that geometry enters only through the operator and coordinate map.

\section{Exact-factorization kinetic-energy diagnostic}
We adopt the exact-factorization gauge~\cite{abedi_exact_2010,hunter_conditionalprobability_1975},
\begin{equation}
  \chi_{\rm EF}(\mathbf R)=
  \left[\int d\mathbf r\,|\Psi(\mathbf r,\mathbf R)|^2\right]^{1/2},
  \qquad
  \Phi_{\mathbf R}(\mathbf r)=
  \frac{\Psi(\mathbf r,\mathbf R)}{\chi_{\rm EF}(\mathbf R)},
\end{equation}
for which $\int d\mathbf r\,|\Phi_{\mathbf R}(\mathbf r)|^2=1$ at each $\mathbf R$.

The gauge-covariant decomposition is
\begin{align}
  K_{\rm BO}^{\rm EF}
   & =
  \frac{1}{2M}\sum_I\int d\mathbf R\,
  \left|[-i\nabla_{\mathbf R_I}+\mathbf A_I(\mathbf R)]
  \chi_{\rm EF}(\mathbf R)\right|^2, \\
  K_{\rm non\mbox{-}BO}^{\rm EF}
   & =
  \frac{1}{2M}\sum_I\int d\mathbf R\,
  |\chi_{\rm EF}(\mathbf R)|^2
  \left[
    \langle\nabla_{\mathbf R_I}\Phi_{\mathbf R}|
    \nabla_{\mathbf R_I}\Phi_{\mathbf R}\rangle_{\mathbf r}
    -
    |\langle\Phi_{\mathbf R}|
    \nabla_{\mathbf R_I}\Phi_{\mathbf R}\rangle_{\mathbf r}|^2
    \right],
\end{align}
where
\begin{equation}
  \mathbf A_I(\mathbf R)=
  \langle\Phi_{\mathbf R}|-i\nabla_{\mathbf R_I}|
  \Phi_{\mathbf R}\rangle_{\mathbf r}.
\end{equation}
The second integrand is the squared norm of the component of $\nabla_{\mathbf R_I}\Phi_{\mathbf R}$ orthogonal to $\Phi_{\mathbf R}$; it is non-negative and invariant under the remaining phase gauge.

We evaluate the fixed-$\mathbf R$ electronic inner products by sampling nuclear snapshots from the optimized NNVMC distribution and then performing electron-only Metropolis sampling. Two independent calculations give EF-gauge kinetic-energy ratios of approximately 3.7\%.
The gauge-invariant total nuclear kinetic energy remains approximately 0.386~Ha. The EF ratio is stable when the numbers of nuclear snapshots and electron walkers are increased. It provides a robust measure of kinetic coupling inside the optimized full wave function; a direct energetic comparison can be obtained from a separate BO-constrained optimization.

\section{Benchmark statistics and autocorrelation}
The benchmark inference used $B=1024$ independent chains and $M=10{,}000$ saved records per chain, with 50 Metropolis proposal steps between saved records. If $E_m^i$ is the local energy for record $m$ in chain $i$, we define
\begin{equation}
  E_m=\frac{1}{B}\sum_{i=1}^{B}E_m^i,\qquad
  \sigma_m^2=\frac{1}{B}\sum_{i=1}^{B}(E_m^i-E_m)^2 .
\end{equation}
Table~I of the main text reports the means of $E_m/N$ and $\sigma_m^2/N$ over the saved-record index $m$. The variance is the width of the local-energy distribution. For either saved-record series $X_m=E_m/N$ or $X_m=\sigma_m^2/N$, we define $\overline X=M^{-1}\sum_{m=1}^{M}X_m$ and $s_X^2=(M-1)^{-1}\sum_{m=1}^{M}(X_m-\overline X)^2$. The parenthesized uncertainties in Table~I of the main text are $\Delta_{\rm uncorr}(\overline X)=s_X/\sqrt{M}$ and do not include an autocorrelation correction.

Autocorrelation was evaluated along $m$, not across the independent-chain index. For a saved-record series $X_m$, we use
\begin{equation}
  g_X=1+2\sum_{k=1}^{K_X}\rho_X(k),\qquad
  \Delta_{\rm corr}(\overline X)=
  \frac{s_X}{\sqrt{M}}\sqrt{g_X}.
\end{equation}
One unit of lag is one saved-record interval, or 50 proposal steps. For $E/N$, $K_E=20$, $\tau_{\rm int,E}\simeq0.82$, $g_E\simeq1.64$, and $\Delta_{\rm corr}(E/N)=6.9\times10^{-6}$~Ha/atom. For $\sigma^2/N$, $K_{\sigma^2}=1$, $\tau_{\rm int,\sigma^2}\simeq0.5$, $g_{\sigma^2}\simeq1$, and $\Delta_{\rm corr}(\sigma^2/N)=4.8\times10^{-5}$~Ha$^2$/atom.

\section{Training Parameters and Training Curves}

Table~\ref{tab:hyperparameters_nvt} lists the hyperparameters for the benchmark calculation of a dynamic BCC lattice at $r_s = 1.31$ in the NVT ensemble, while Table~\ref{tab:hyperparameters_npt} details those for various initial structures in the broken symmetry phase under the NPT ensemble. For the NPT training, we adopted a multi-stage strategy comprising three parts: (1) Warmup, where lattice parameters are fixed and wavefunction parameters are optimized to establish a high-quality initial wavefunction; (2) Lattice training, where wavefunction parameters are frozen while lattice parameters are optimized; and (3) Wavefunction training, which mirrors the warmup phase but involves fewer steps. The overall protocol begins with the initial Warmup phase, followed by an alternating cycle of Lattice training and Wavefunction training.

\begin{figure}[htbp]
  \centering
  \includegraphics[width=0.46\textwidth]{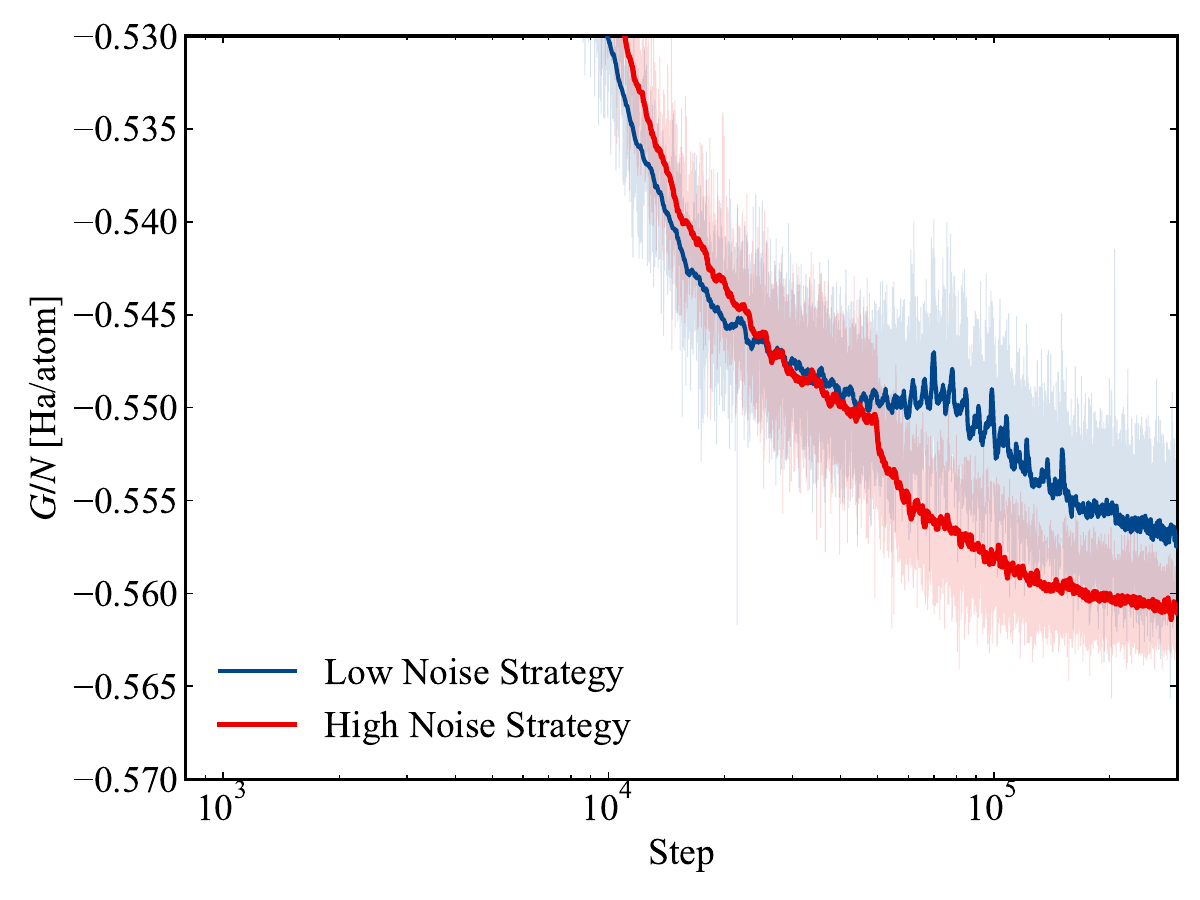}
  \includegraphics[width=0.46\textwidth]{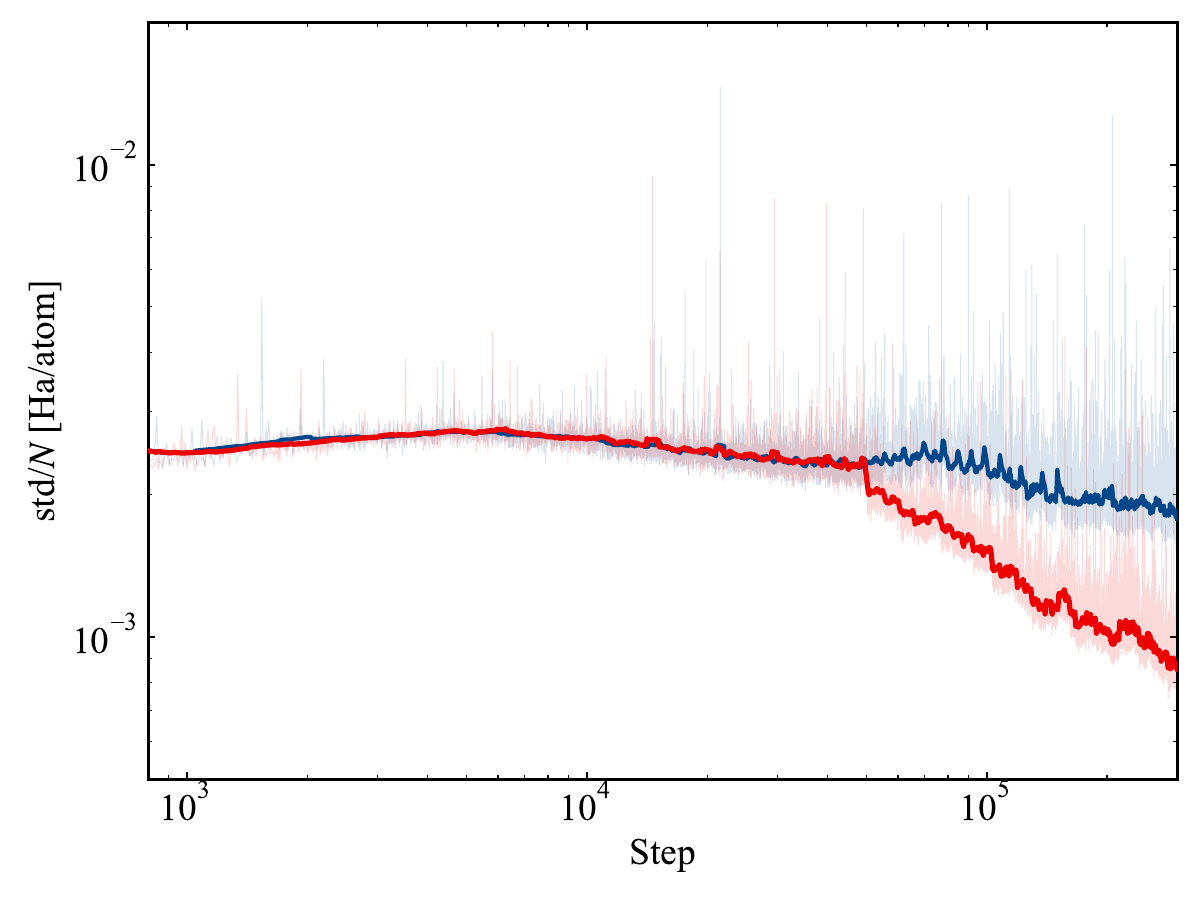}
  \caption{Comparison of different annealing strategies. The high-noise strategy (single estimate, more steps) demonstrates faster convergence and lower variance compared to the low-noise strategy (averaged estimates), despite identical computational costs. The shadow in the figure represents the original data points, while the solid line is the smoothed curve via a moving average with a window size of 1000 steps.
  }
  \label{fig:train_curve_Annealing}
\end{figure}

To demonstrate that noise may be beneficial in the simulated annealing process, Fig.~\ref{fig:train_curve_Annealing} compares the training curves of two distinct annealing strategies: (i) a low-noise scheme that averages 10 independent enthalpy estimates per step, and (ii) a high-noise scheme that uses a single estimate but allows for 10 times as many steps. By design, the total computational cost for enthalpy evaluation remains identical for both methods. The results indicate that while both strategies converge to approximately the same enthalpy, the high-noise strategy exhibits faster convergence and yields lower variance. This demonstrates that the statistical noise of the enthalpy estimator supplies an effective stochastic driving force that enhances barrier crossing and accelerates convergence.

\begin{figure}[htbp]
  \centering
  \includegraphics[width=0.46\textwidth]{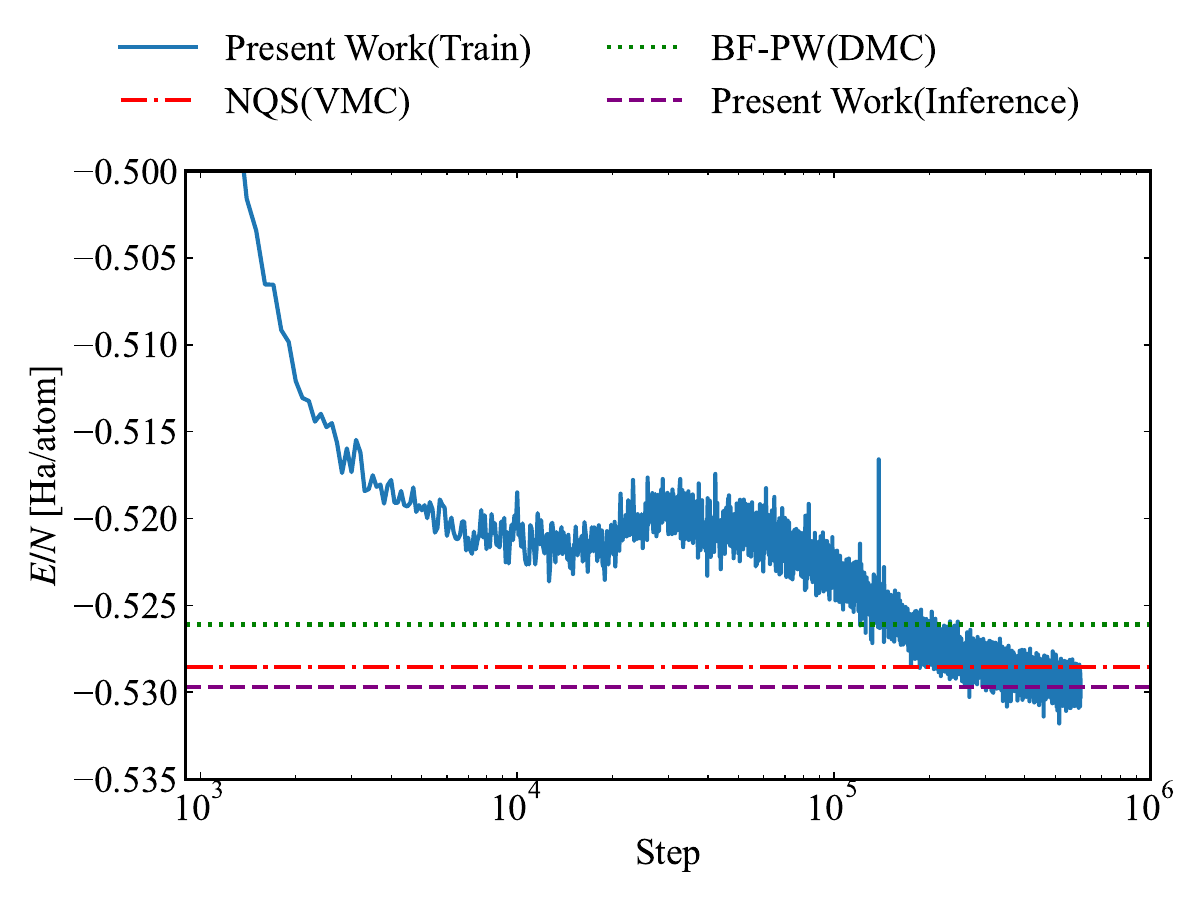}
  \includegraphics[width=0.46\textwidth]{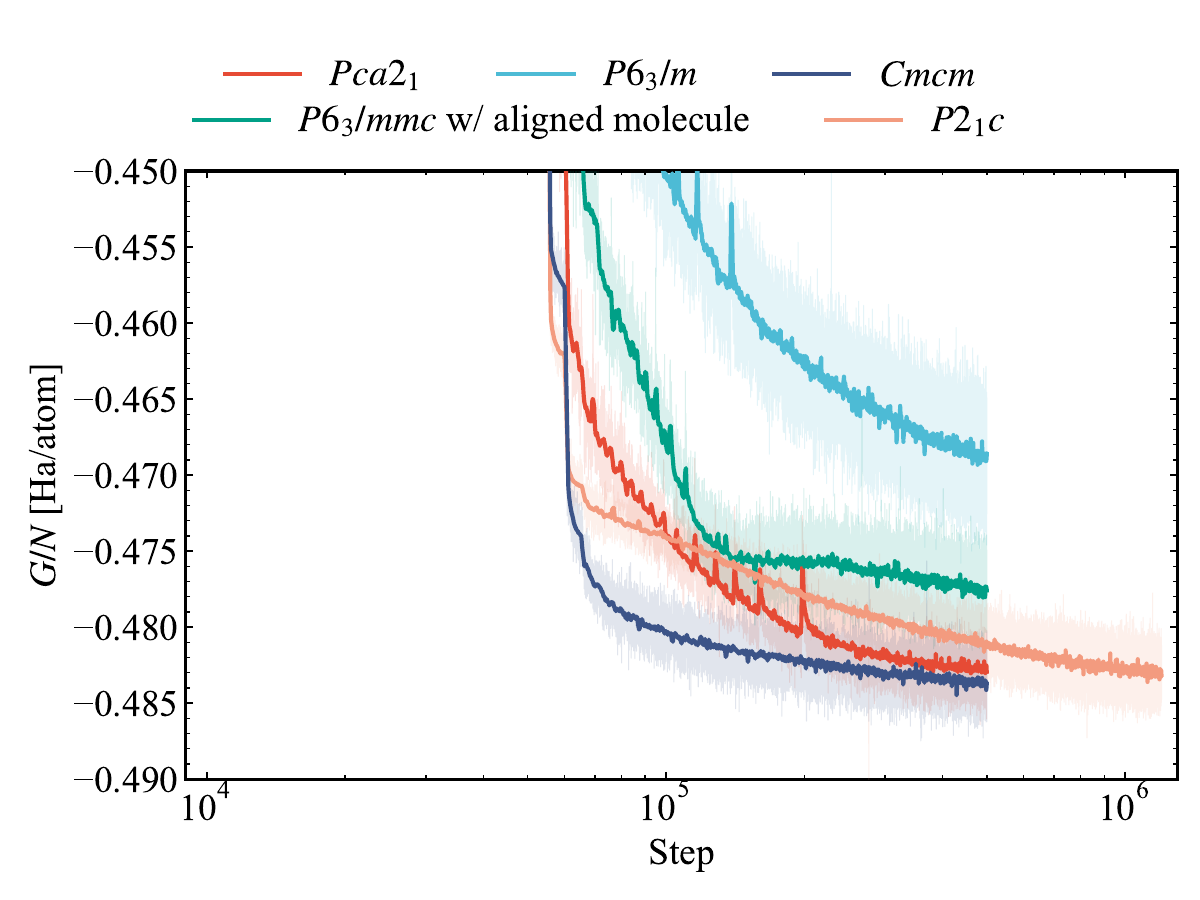}
  \caption{\textbf{Left}: Training curves for the BCC dynamic lattice at $r_s = 1.31$ in the NVT ensemble.
    \textbf{Right}: Training curves for various initial structures at $P_{\text{ext}} = 130$ GPa in the NPT ensemble, highlighting the stability of the $Cmcm$ structure. The legend denotes the initial structure. The shadow in the figure represents the original data points, while the solid line is the smoothed curve via a moving average with a window size of 1000 steps.
  }
  \label{fig:train_curve_NVT_NPT}
\end{figure}

\begin{figure}[htbp]
  \centering
  \includegraphics[width=0.46\textwidth]{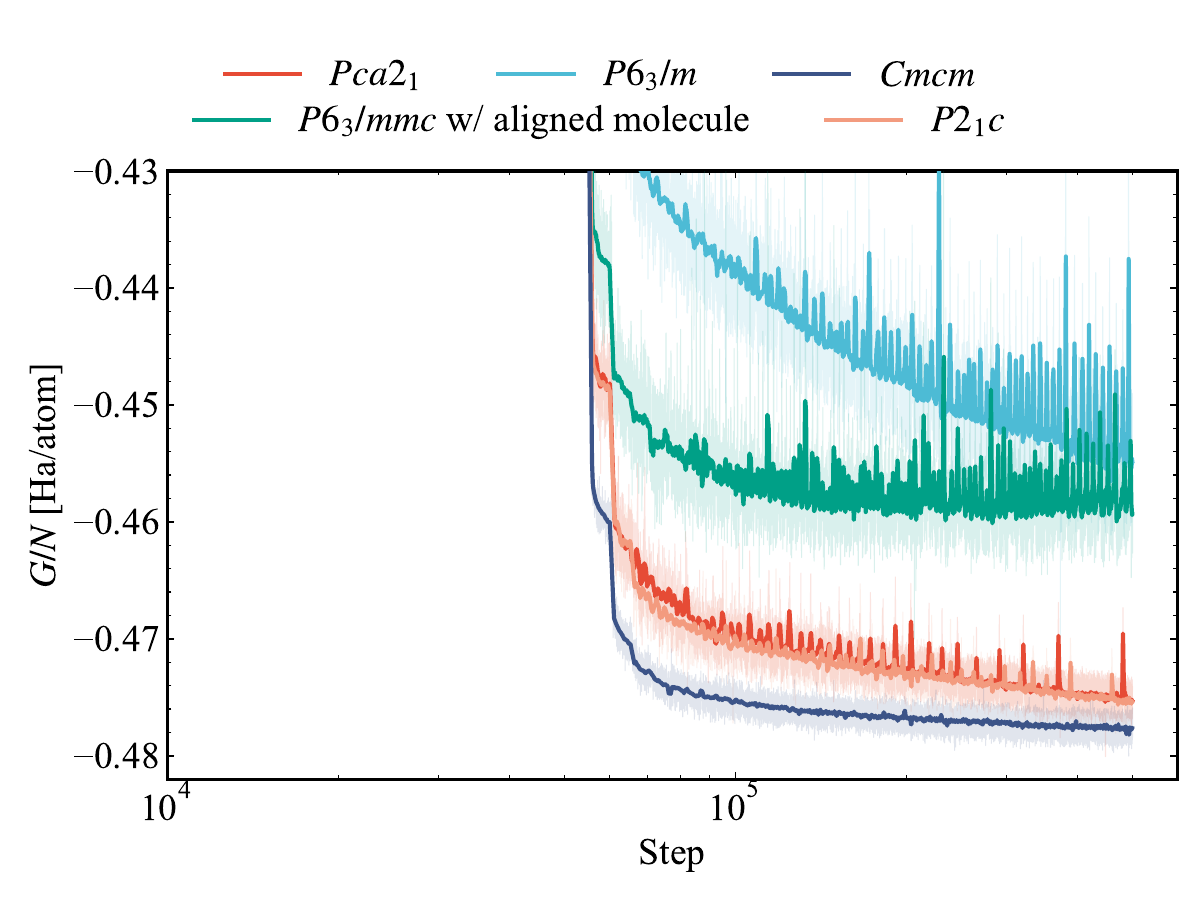}
  \caption{Training curves for various initial structures at $P_{\text{ext}} = 130$~GPa in the NPT ensemble with $2\times 2\times 2$ twist averaged boundary conditions. The legend denotes the initial structure. Shaded regions represent the original data points, while solid lines show smoothed curves obtained via a moving average with a window size of 1000 steps.}
  \label{fig:train_curve_NPT_twist}
\end{figure}

In Fig.~\ref{fig:train_curve_NVT_NPT}, the left panel displays the training curves for the BCC dynamic lattice at $r_s = 1.31$ within the NVT ensemble. Here, we benchmark our approach against other VMC methods, including NQS and traditional DMC, the latter of which serves as the lowest-energy reference in traditional methods. The right panel illustrates the training curves for various initial structures at $P_{\text{ext}} = 130$ GPa within the NPT ensemble($P2_1/c$ is first training at $P_\text{ext}=10$ GPa, and then at 130 GPa). These results identify the $Cmcm$ phase as the lowest-energy structure among the tested structures at this pressure. The final enthalpies based on inference with 10000 steps after training are summarized in Table~\ref{tab:final_enthalpies}. The structure with the lowest enthalpy is listed in Table~\ref{tab:cmcm_structure}. To assess finite size effects, we also performed twist averaged boundary conditions (TABC)~\cite{Lin:2001zz} calculations with a $2\times 2\times 2$ twist grid; the training curves are shown in Fig.~\ref{fig:train_curve_NPT_twist} and the corresponding enthalpies are compared alongside the PBC results in Table~\ref{tab:final_enthalpies}.

\begin{figure}[htbp]
  \centering
  \includegraphics[width=0.9\textwidth]{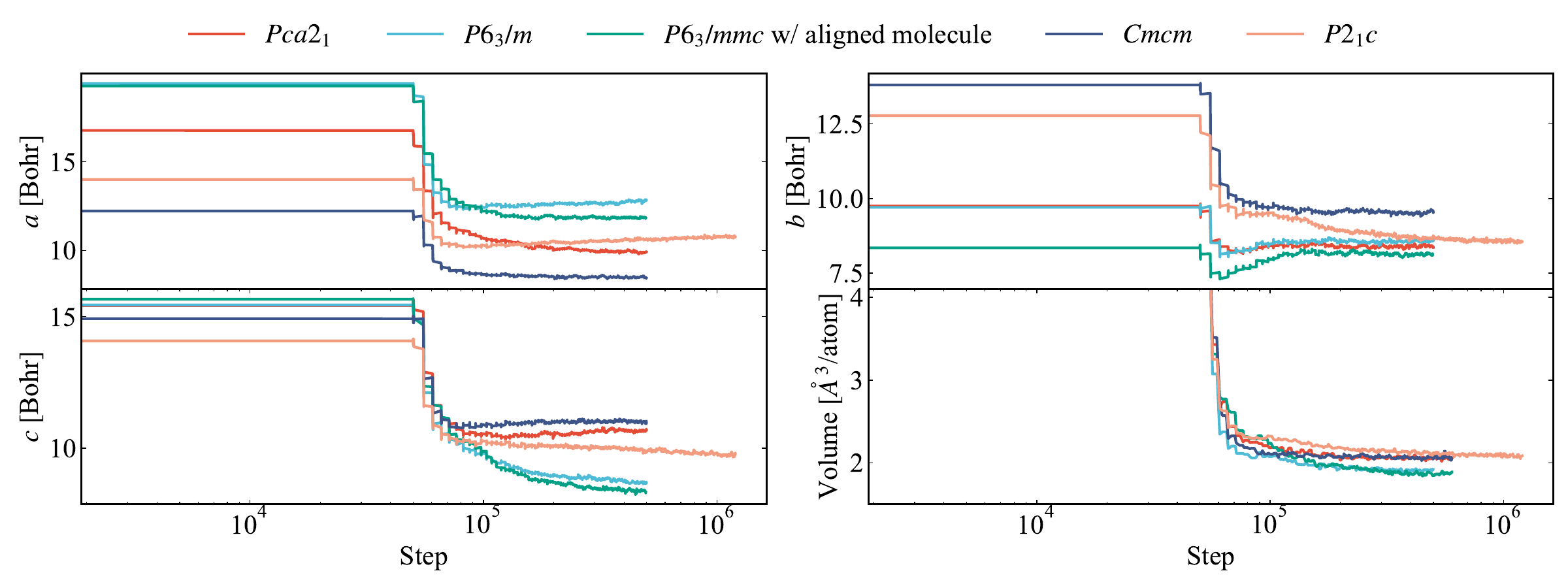}
  \caption{The evolution of lattice parameters and volume per atoms as a function of training steps. The legend denotes the initial structure.}
  \label{fig:train_curve_lattice}
\end{figure}

\begin{figure}[htbp]
  \centering
  \includegraphics[width=\textwidth]{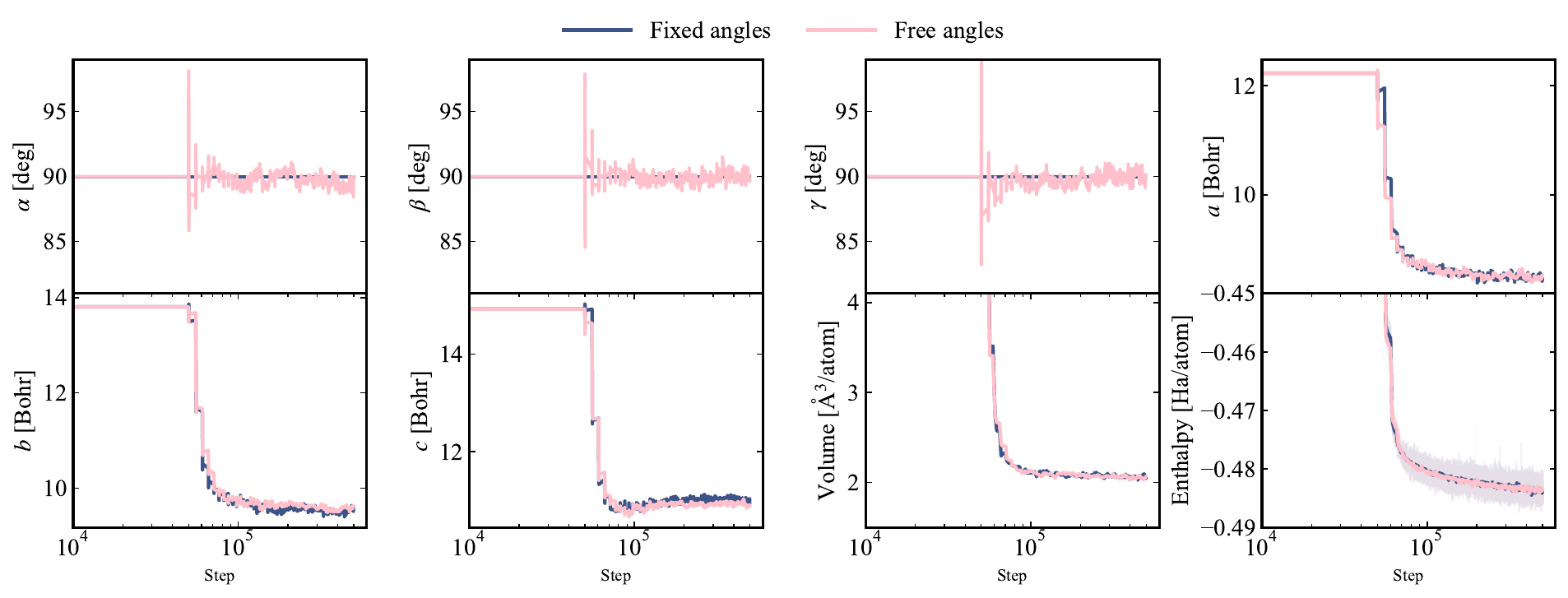}
  \caption{Comparison between two NPT training runs starting from a $Cmcm$ structure with 64 atoms at 130 GPa: one with cell angles constrained to $\alpha=\beta=\gamma=90^\circ$ ('Fixed angles') and another with angles freely optimized along with other lattice parameters ('Free angles'). Shaded regions in enthalpy curves represent the original data points, while solid lines show smoothed curves obtained via a moving average with a window size of 1000 steps.}
  \label{fig:free_angle}
\end{figure}

In Fig.~\ref{fig:train_curve_lattice}, we present the evolution of the lattice parameters and the volume per atom as a function of the training steps. For simplicity, the lattice angles were fixed at $\alpha=\beta=\gamma=90^{\circ}$ in Eq.~\ref{eq:lattice} throughout the training, and allowing them to relax produces no appreciable deviation from this constraint as shown in Fig.~\ref{fig:free_angle}.

\begin{table}[htbp]
  \centering
  \caption{Hyperparameters for BCC Train}
  \label{tab:hyperparameters_nvt}
  \begin{tabular}{cc|cc}
    \toprule
    Hyperparameter                  & Value & Hyperparameter                  & Value   \\
    \midrule
    Dimension of one-electron layer & 256   & Dimension of two-electron layer & 32      \\
    Dimension of nuclear layer      & 32    & Number of layers                & 3       \\
    Number of determinants          & 16    & Optimizer                       & KFAC    \\
    Learning rate                   & 3e-2  & Learning rate  decay            & 1       \\
    Learning rate delay             & 1e4   & Damping                         & 1e-3    \\
    Momentum of optimizer           & 0.0   & Batch size                      & 1024    \\
    Number of training steps        & 6e5   & GBS burn in                     & 100     \\
    Number of GBS blocks            & 5     & Number of GBS iterations        & 10      \\
    Electron move width             & 2e-2  & Nuclear move width              & 2e-4    \\
    Adapt frequency                 & 100   & Precision                       & Float32 \\
    \bottomrule
  \end{tabular}
\end{table}
\begin{table}[htbp]
  \centering
  \caption{Hyperparameters for Broken Symmetry Phase}
  \label{tab:hyperparameters_npt}
  \begin{tabular}{cc|cc}
    \toprule
    Hyperparameter                                                                        & Value & Hyperparameter                  & Value   \\
    \midrule
    Dimension of one-electron layer                                                       & 256   & Dimension of two-electron layer & 32      \\
    Dimension of nuclear layer                                                            & 32    & Number of layers                & 3       \\
    Number of determinants                                                                & 8     & Optimizer                       & KFAC    \\
    Learning rate                                                                         & 3e-3  & Learning rate  decay            & 1       \\
    Learning rate delay                                                                   & 1e4   & Damping                         & 1e-3    \\
    Momentum of optimizer                                                                 & 0.0   & Batch size                      & 1024    \\
    Number of training steps\footnote{For $P2_1c$, the number of training steps is 1.2e6} & 5e5   & GBS burn in                     & 100     \\
    Number of GBS blocks                                                                  & 10    & Number of GBS iterations        & 10      \\
    Electron move width                                                                   & 2e-2  & Nuclear move width              & 2e-4    \\
    Adapt frequency                                                                       & 100   & Precision                       & Float32 \\
    Beta in Annealing                                                                     & 0.002 & Initial virtual temperature     & 0.1     \\
    Annealing steps between each training steps                                           & 10    & Lattice move width              & 0.02    \\
    Warmup steps                                                                          & 50000 & Lattice steps                   & 300     \\
    Wavefunction steps                                                                    & 5000  & Initial $r_s$                   & 2.11    \\
    \toprule
  \end{tabular}
\end{table}


\begin{table}[htbp]
  \centering
  \caption{Final enthalpies based on inference with 10000 steps after training, including PBC and TABC.}
  \label{tab:final_enthalpies}
  \begin{tabular}{ccc}
    \toprule
    Initial Structure                 & \multicolumn{2}{c}{Final Enthalpy (Ha/atom)}                         \\
    \cmidrule(lr){2-3}
                                      & PBC                                          & TABC                  \\
    \midrule
    $Cmcm$                            & \textbf{-0.483666(10)}                       & \textbf{-0.477739(6)} \\
    $Pca2_1$                          & -0.482817(9)                                 & -0.475292(6)          \\
    $P2_1c$                           & -0.482795(9)                                 & -0.475416(6)          \\
    $P6_3/mmc$  with aligned molecule & -0.47791(1)                                  & -0.45908(1)           \\
    $P6_3m$                           & -0.46987(2)                                  & -0.45477(2)           \\
    \bottomrule
  \end{tabular}
\end{table}

\begin{table}[htbp]
  \centering
  \caption{The structure identified at 130~GPa for the broken symmetry phase.}
  \label{tab:cmcm_structure}
  \begin{tabular}{l c c c c c c}
    \toprule
    \multirow{2}{*}{Space group} & \multirow{2}{*}{Lattice parameters (Bohr, deg)} &
    \multirow{2}{*}{Wyckoff position}
                                 & \multicolumn{3}{c}{Fractional coordinates}
    \\ & & & $x$   & $y$    & $z$ \\
    \midrule
    \multirow{4}{*}{$Cmcm$}
                                 & $a = 8.48$                                      & 8g     & 0.5992  & 0.3967 & 0.2500 \\
                                 & $b = 9.60$                                      & 8g     & -0.0825 & 0.2402 & 0.2500 \\
                                 & $c = 5.47$                                      & 8g     & 0.7642  & 0.3967 & 0.2500 \\
                                 & $\alpha=\beta=\gamma=90.0$
                                 & 8g                                              & 0.4175 & 0.0867  & 0.2500          \\
    \bottomrule
  \end{tabular}
\end{table}

\section{XRD analysis of candidate structures}
\begin{figure}[htbp]
  \centering
  \includegraphics[width=0.9\textwidth]{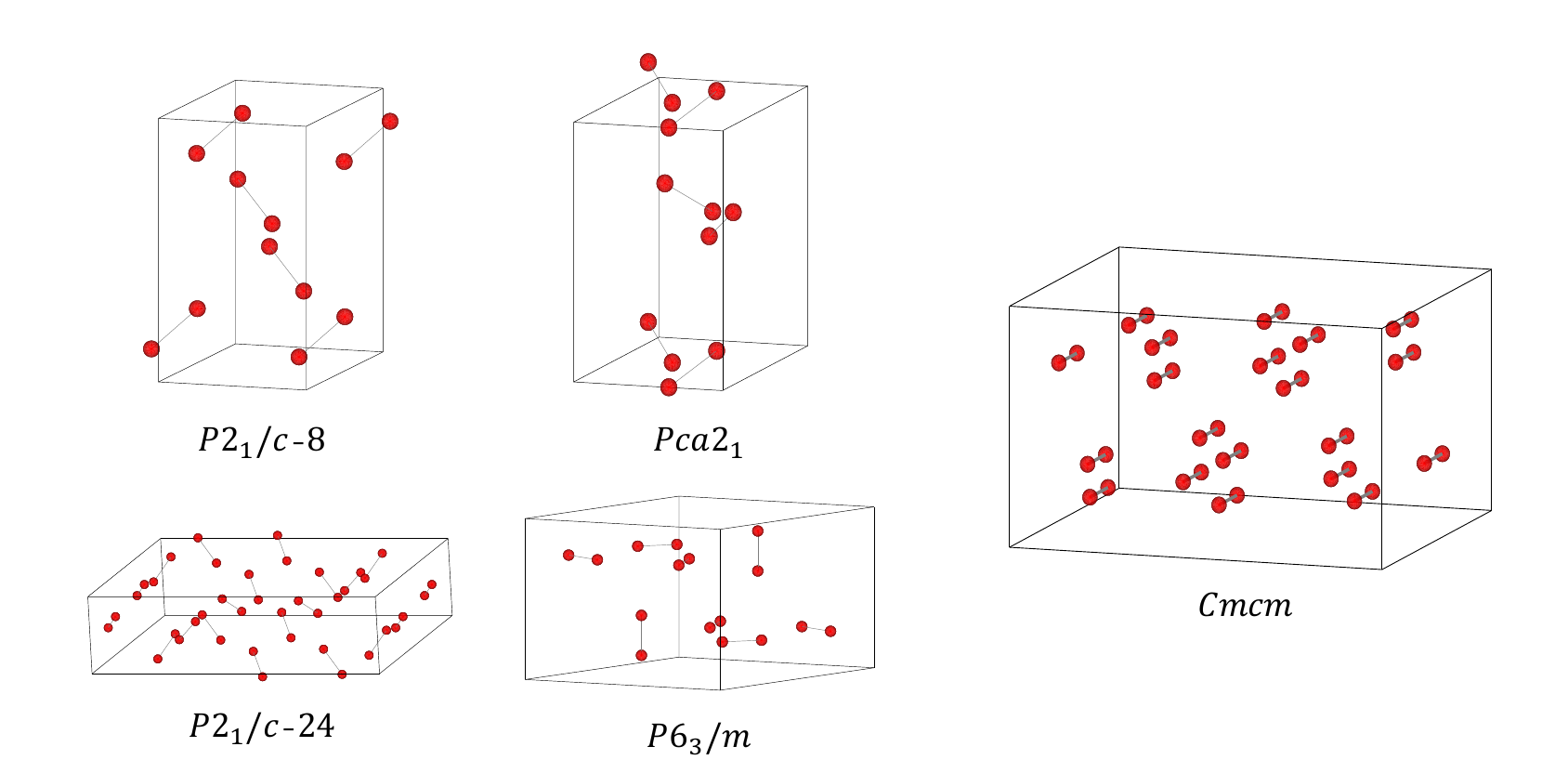}
  \caption{The structures used in the XRD patterns in Fig.~4 of the main text. The solid line represents the bonds between the hydrogen atoms, while the box represents the conventional cell for $Cmcm$ and primitive cell for other structures.}
  \label{fig:structure_xrd}
\end{figure}

Figure~\ref{fig:structure_xrd} displays the structures analyzed in the XRD patterns of Fig.~4 of the main text. The $P2_1/c\text{-}8$ and $P2_1/c\text{-}24$ phases were taken from Refs.~\cite{li_highpressurestructures_2024} and \cite{pickard_structures_2009}, respectively, with all other structures taken from Ref.~\cite{pickard_structurephase_2007}. A DFT variable-cell relaxation was performed to obtain the structural parameters at 130~GPa.

For each candidate, the common scale factor is $s=(V_{\rm exp}/V_{\rm calc})^{1/3}$, so that $d_{hkl}^{\rm scaled}=s\,d_{hkl}^{\rm calc}$. We quantify the remaining peak-position differences with
\begin{align}
  \Delta_{\rm sim\to exp}
   & =
  \left[\frac{1}{N_{\rm sim}}\sum_i\min_j
    (d_i^{\rm sim}-d_j^{\rm exp})^2\right]^{1/2}, \\
  \Delta_{\rm exp\to sim}
   & =
  \left[\frac{1}{N_{\rm exp}}\sum_j\min_i
    (d_j^{\rm exp}-d_i^{\rm sim})^2\right]^{1/2}.
\end{align}
These metrics provide a reproducible peak-position comparison complementary to a full-profile structural refinement; the results are summarized in Table~\ref{tab:xrd_peak_position_comparison}.

\begin{table}[htbp]
  \centering
  \caption{RMS peak-position comparison after uniform volume rescaling. The left block includes all simulated peaks above the common intensity threshold; the right block is restricted to $1.35\le d\le1.65$~\AA. All RMS differences are in \AA.}
  \label{tab:xrd_peak_position_comparison}
  \footnotesize
  \setlength{\tabcolsep}{3pt}
  \begin{tabular}{@{}lcccccc@{}}
    \toprule
                       & \multicolumn{3}{c}{All selected peaks}
                       & \multicolumn{3}{c}{Near experimental peaks}                                                                     \\
    \cmidrule(lr){2-4}\cmidrule(lr){5-7}
    Candidate          & $N_{\rm all}$                               & $\Delta_{\rm sim\to exp}$
                       & $\Delta_{\rm exp\to sim}$                   & $N_{\rm near}$
                       & $\Delta_{\rm sim\to exp}$
                       & $\Delta_{\rm exp\to sim}$                                                                                       \\
    \midrule
    $Cmcm$             & 7                                           & \textbf{0.1248}           & \textbf{0.0036}
                       & 4                                           & \textbf{0.0039}           & \textbf{0.0036}                       \\
    $P2_1/c\text{-}8$  & 9                                           & 0.1482                    & 0.0116          & 6 & 0.0178 & 0.0116 \\
    $P2_1/c\text{-}24$ & 12                                          & 0.1447                    & 0.0096          & 7 & 0.0286 & 0.0096 \\
    $Pca2_1$           & 8                                           & 0.1570                    & 0.0078          & 5 & 0.0089 & 0.0078 \\
    $P6_3/m$           & 8                                           & 0.1603                    & 0.0078          & 4 & 0.0220 & 0.0078 \\
    \bottomrule
  \end{tabular}
\end{table}

To estimate coherent Bragg-intensity changes from the sampled nuclear density, we first remove global center-of-mass drift by aligning every $|\Psi|^2$ nuclear configuration to the static reference and wrapping it into the periodic cell. For reciprocal vector $\mathbf G_\alpha=2\pi L^{-T}\mathbf h_\alpha$, define
\begin{equation}
  \rho_s(\mathbf G)=\sum_{j=1}^{N_{\rm atom}}
  e^{i\mathbf G\cdot\mathbf r_{s,j}},\qquad
  \rho_{\rm static}(\mathbf G)=\sum_j
  e^{i\mathbf G\cdot\mathbf r_j^{\rm static}} .
\end{equation}
For peak group $p$, with multiplicity $m_\alpha$,
\begin{align}
  S_p^{\rm static}
   & =\sum_{\alpha\in p}m_\alpha
  |\rho_{\rm static}(\mathbf G_\alpha)|^2, \\
  S_p^{\rm coh}
   & =\sum_{\alpha\in p}m_\alpha
  |\langle\rho_s(\mathbf G_\alpha)\rangle_s|^2,
\end{align}
and the coherent correction is
\begin{equation}
  R_p^{\rm coh}=\frac{S_p^{\rm coh}}{S_p^{\rm static}},\qquad
  I_p^{\rm coh,corr}=I_p^{\rm static}R_p^{\rm coh}.
\end{equation}
This construction retains the static peak positions and X-ray prefactors while incorporating the coherent Bragg-intensity correction. A full quantum diffraction treatment would additionally describe diffuse scattering and peak-position shifts. The correction suppresses the feature near the first experimental line and the weak features near 1.3 and 1.7~\AA{}, while slightly enhancing the third feature. These intensity trends provide a qualitative view of quantum nuclear effects, with the third feature remaining weak.

\section{Raman and IR analysis}
This analysis uses the correlation method~\cite{nakamoto_infrared_2009} to determine the vibrational properties of solid hydrogen. The isolated $\mathrm{H}_2$ molecule belongs to the linear $D_{\infty h}$ point group. In the solid state, hydrogen crystallizes in the orthorhombic space group $Cmcm$, which corresponds to the factor group $D_{2h}$.The molecular centers of mass occupy three distinct Wyckoff positions: two $4c$ sites and one $8g$ site. The site symmetries for these molecular centers are $C_{2v}$ for the $4c$ positions and $C_s$ for the $8g$ positions. The derivation of the Raman and IR active modes proceeds by:
\begin{enumerate}
  \item characterizing the modes of the isolated $\mathrm{H}_2$ molecule;
  \item constructing correlation tables linking $D_{\infty h}$ to the site symmetries $C_{2v}$ and $C_s$;
  \item correlating these site symmetries with the factor group $D_{2h}$; and
  \item deriving the selection rules for the crystal lattice.
\end{enumerate}
\subsection{Mode of Hydrogen Molecule $\mathrm{H}_2$}
A diatomic molecule possesses 6 degrees of freedom (dof):
\begin{enumerate}
  \item Translations (3 dof): Basis functions $x, y, z$, corresponding to $\Pi_u(x, y)$ and $\Sigma_u^+(z)$ representations.
  \item Rotations (2 dof): Basis functions $R_x, R_y$, transforming as $\Pi_g(R_x, R_y)$.
  \item Vibration (1 dof): The vibrational mode transforms as $\Sigma_g^+$.
\end{enumerate}
\subsection{Correlation table for $D_{\infty h}$ to site symmetry $C_{2v}$ and $C_s$}
We first restrict the representation in $D_{\infty h}$ into site symmetries $C_{2v}$ and $C_s$, it can be established from standard correlation relation table.
\begin{enumerate}
  \item $D_{\infty h} \rightarrow C_{2v}$:
        \begin{enumerate}
          \item$\Sigma_g^+\to A_g\to A_{1}$
          \item $\Pi_g\to B_{2g}+B_{3g}\to A_{2}+B_{1}$
          \item $\Pi_u\to B_{2u}+B_{3u}\to A_{1}+B_{2}$
          \item $\Sigma_u^+\to B_{1u}\to B_{1}$
        \end{enumerate}
  \item $D_{\infty h} \rightarrow C_s$:
        \begin{enumerate}
          \item $\Sigma_g^+\to A_g\to A^{'}$
          \item $\Pi_g\to B_{2g}+B_{3g}\to 2A^{''}$
          \item $\Pi_u\to B_{2u}+B_{3u}\to 2A^{'}$
          \item $\Sigma_u^+\to B_{1u}\to A^{''}$
        \end{enumerate}
\end{enumerate}
\subsection{Induced Representations in the Factor Group $D_{2h}$}
To determine the representations induced from the subgroups $C_{2v}$ and $C_s$ into the factor group $D_{2h}$, we employ Frobenius reciprocity. This principle states that the multiplicity of an irreducible representation $\rho$ of a group $G$ in the induced representation of a subgroup $H$ is equal to the multiplicity of the representation of $H$ in the restriction of $\rho$ to $H$, $  \langle \mathrm{Ind}_H^G(\sigma),\, \rho \rangle_G
  =
  \langle \sigma,\, \mathrm{Res}_H(\rho) \rangle_H$.
Consequently, the standard correlation tables from $D_{2h}$ down to $C_{2v}$ and $C_s$ allow us to determine the induced modes.

\begin{enumerate}
  \item $C_{2v} \rightarrow D_{2h}$:
        \begin{enumerate}
          \item $A_{1}\to A_{g}+B_{2u}$
          \item $A_{2}\to A_{u}+B_{2g}$
          \item $B_{1} \to B_{3g}+B_{1u}$
          \item $B_{2} \to B_{1g}+B_{3u}$
        \end{enumerate}
  \item $C_s \rightarrow D_{2h}$:
        \begin{enumerate}
          \item $A^{'}\to A_{g}+B_{1g}+B_{2u}+B_{3u}$
          \item $A^{''}\to A_{u}+B_{1u}+B_{2g}+B_{3g}$
        \end{enumerate}
\end{enumerate}
\subsection{Raman and IR Active Modes of Solid Hydrogen}
Thus, for the vibron:
\begin{enumerate}
  \item site A($4c$): $\Sigma_g^+\to A_g\to A_{1}\to A_{g}+B_{2u}$
  \item site B($4c$): $\Sigma_g^+\to A_g\to A_{1}\to A_{g}+B_{2u}$
  \item site C($8g$): $\Sigma_g^+\to A_g\to A^{'}\to A_{g}+B_{1g}+B_{2u}+B_{3u}$
\end{enumerate}
Thus, the representation of the vibron is
\begin{equation}
  \Gamma_{vibron} = 2(A_{g}+B_{2u})+(A_{g}+B_{1g}+B_{2u}+B_{3u})=3A_{g}+3B_{2u}+B_{1g}+B_{3u}
\end{equation}

For the Libron(phonon):
\begin{enumerate}
  \item site A($4c$): $\Pi_g\to B_{2g}+B_{3g}\to A_{2}+B_{1}\to A_{u}+B_{2g}+B_{3g}+B_{1u}$
  \item site B($4c$): $\Pi_g\to B_{2g}+B_{3g}\to A_{2}+B_{1}\to A_{u}+B_{2g}+B_{3g}+B_{1u}$
  \item site C($8g$): $\Pi_g\to B_{2g}+B_{3g}\to 2A^{''}\to 2A_{u}+2B_{1u}+2B_{2g}+2B_{3g}$
\end{enumerate}
Thus, the representation of the libron is
\begin{equation}
  \begin{aligned}
    \Gamma_{libron} & = 2(A_{u}+B_{2g}+B_{3g}+B_{1u})+2A_{u}+2B_{1u}+2B_{2g}+2B_{3g} \\
                    & = 4A_{u}+4B_{1u}+4B_{2g}+4B_{3g}
  \end{aligned}
\end{equation}
Translational Phonons
\begin{enumerate}
  \item site A($4c$): $\Sigma_u^++\Pi_u\to B_{1u} +B_{2u}+B_{3u} \to A_{1}+B_{1}+B_{2}\to A_{g}+B_{2u}+B_{3g}+B_{1u}+B_{1g}+B_{3u}$
  \item site B($4c$): $\Sigma_u^++\Pi_u\to B_{1u} +B_{2u}+B_{3u} \to A_{1}+B_{1}+B_{2}\to A_{g}+B_{2u}+B_{3g}+B_{1u}+B_{1g}+B_{3u}$
  \item site C($8g$): $\Sigma_u^++\Pi_u\to B_{1u} +B_{2u}+B_{3u}\to 2A^{'}+A^{''}\to A_{u}+B_{1u}+B_{2g}+B_{3g}+2A_{g}+2B_{1g}+2B_{2u}+2B_{3u}$
\end{enumerate}
Summing over all sites yields:
\begin{equation}
  \begin{aligned}
    \Gamma_{trans} & = 2A_{g}+2B_{2u}+2B_{3g}+2B_{1u}+2B_{1g}+2B_{3u}              \\
                   & +A_{u}+B_{1u}+B_{2g}+B_{3g}+2A_{g}+2B_{1g}+2B_{2u}+2B_{3u}    \\
                   & = A_{u}+4B_{1g}+4B_{2u}+3B_{3g}+4A_{g}+B_{2g}+3B_{1u}+4B_{3u}
  \end{aligned}
\end{equation}
Subtracting the acoustic branches ($B_{1u} + B_{2u} + B_{3u}$):
\begin{equation}
  \begin{aligned}
    \Gamma_{optical} & =  \Gamma_{trans}- \Gamma_{acoustic}                           \\
                     & =  A_{u}+4B_{1g}+3B_{2u}+3B_{3g}+4A_{g}+B_{2g}+2B_{1u}+3B_{3u}
  \end{aligned}
\end{equation}
The optical phonon is
\begin{equation}
  \Gamma_{optpho} = \Gamma_{optical}+\Gamma_{libron} = 5A_{u}+4B_{1g}+3B_{2u}+7B_{3g}+4A_{g}+5B_{2g}+6B_{1u}+3B_{3u}
\end{equation}
Note that $A_u$ modes are silent (inactive in both Raman and IR). In summary, the Raman and IR active modes are:
\begin{align}
  \Gamma_{\mathrm{Raman}}^{\mathrm{vibron}} & = 3A_g + B_{1g},                      \\
  \Gamma_{\mathrm{IR}}^{\mathrm{vibron}}    & = 3B_{2u} + B_{3u},                   \\
  \Gamma_{\mathrm{Raman}}^{\mathrm{phonon}} & = 4B_{1g} + 7B_{3g} + 4A_g + 5B_{2g}, \\
  \Gamma_{\mathrm{IR}}^{\mathrm{phonon}}    & = 3B_{2u} + 6B_{1u} + 3B_{3u}.
\end{align}

These correlation tables establish the symmetry-allowed modes. Comparison with the observed subset therefore provides a symmetry-level compatibility check, while frequencies and intensities remain targets for future calculations.

\end{document}